\documentclass[useAMS,usenatbib,usegraphicx]{mn2e}
\usepackage{amssymb}

\usepackage{times} 

\newcommand{\degree}{^\circ}

\def \<{\langle}
\def \>{\rangle}
\def\acknowledgments {\section*{Acknowledgements}%
  \addtocontents{toc}{\protect\vspace{6pt}}%
  \addcontentsline{toc}{section}{Acknowledgements}%
}

\title[Improved CMB Map from WMAP Data]
{Improved CMB Map from WMAP Data}

\author[H. Liu \& T. P. Li]{Hao Liu$^{1}$\thanks{E-mail: liuhao@ihep.ac.cn}
and Ti-Pei Li$^{1,2}$\thanks{E-mail: litp@tsinghua.edu.cn}\\
$^{1}$Key Laboratory of Particle Astrophysics, Institute of High Energy Physics,
Chinese Academy of Sciences, Beijing, China\\
$^2$Center for Astrophysics, Tsinghua University, Beijing, China
}
\begin{document}

\date{}

\pagerange{\pageref{firstpage}--\pageref{lastpage}} \pubyear{2009}

\maketitle

\label{firstpage}

\begin{abstract}
The cosmic microwave background
(CMB) temperature maps published by the Wilkinson Microwave
Anisotropy Probe (WMAP) team are found to be inconsistent with the differential
time-ordered data (TOD), from which the maps are reconstructed.
The inconsistency indicates that there is a serious problem in the map making routine
of the WMAP team, and it is necessary to reprocess the WMAP data.
We develop a self-consistent software package of map-making and power
spectrum estimation independently of the WMAP team.
Our software passes a variety of tests. New
CMB maps are then reconstructed, which are significantly different from the
official WMAP maps. In the new maps, the inconsistency disappeared,
along with the hitherto unexplained high level alignment between
the CMB quadrupole and octopole components detected in released WMAP
maps.
An improved CMB cross-power spectrum is then
derived from the new maps which better agrees with that of
BOOMRANG.  Two important results are hence obtained:
the CMB quadrupole drops to nearly
zero, and the power in multiple moment range between 200 and 675
decreases on average by about $13\%$, causing the best-fit
cosmological parameters to change considerably, e.g., the total matter
density increases from 0.26 up to 0.32 and the dark energy density
decreases from 0.74 down to 0.68. These new parameters match
with improved accuracy those of other independent experiments.
Our results indicate that there is still room for significant revision
in the cosmological model parameters.
\end{abstract}


\section{INTRODUCTION}
The CMB data from the WMAP mission is the most important basis of
cosmology study, and the accuracies of the CMB map recovered from
the WMAP data and its angular power spectrum are essential for precision
cosmology.

Recently, we find there notably exist observational effects on
released WMAP maps. The WMAP mission measures temperature
differences between sky points using differential radiometers
consisting of plus-horn and minus-horn~\cite{ben03a}. When an
antenna horn points to a sky pixel, the other one will scan a ring
in the sky with angular radius $141\degree$ to the center pixel.
These measured TOD are transformed into the full-sky
temperature anisotropy map by a map-making
process~\cite{hin03a}. In released five-year WMAP (WMAP5) CMB maps
we find significant distortion from hot Galactic sources: the pixels
in the scan ring of a hot pixel are systematically cooled, and
strongest anti-correlations between temperatures of a hot pixel and
its scan-ring appear at a separation angle
$\theta\sim141\degree$~\cite{liu09}. The above results are confirmed
by Aurich, Lustig \& Steiner (2009). Furthermore, we also detect the
no-negligible effect of imbalance observations in published WMAP5
maps~\cite{li09}:
 systematical dependence of temperature vs. observation number difference between
the two horns of a radiometer produced by the input transmission imbalance,
and significant correlation between pixel temperature
and observation number in WMAP data. These recent findings of systematical error
in published WMAP temperature maps push us forward to further check
the WMAP map-making processing.

In this work, we find a remarkable inconsistency between the WMAP
TOD and published temperature map, which is described in \S2.
The revealed inconsistency demonstrates that there certainly exists
a serious problem in WMAP map-making process, and it is worth
to check the reliability of released WMAP results by reproducing CMB temperature
maps from the original raw data independently from the WMAP team.
We built a self-consistent software package of map-making from TOD and power
spectrum estimation from reconstructed CMB maps.
Our software successfully passes a variety of tests, i.e. the residual TOD test,
the residual dipole component test, the map-making
convergence test and the end-to-end test, which
are presented in \S3. With our software, improved CMB maps from WMAP TOD
are produced and shown in \S4, and the new angular power spectrum is derived from
our new CMB maps and shown in \S5. From the new angular power spectrum, we determine
the new best-fit cosmological parameters
and present the results in \S6. Finally, we give a brief discussion of our results
in \S7.

\section{INCONSISTENCY BETWEEN TOD AND RELEASED MAP}
Let  $t_i$ denotes the temperature anisotropy at a sky pixel $i$.
In a certain band, the observed difference of the $k$th observation
$ d_k = t_{k^+} - t_{k^-}$,
where $k^+$ and $k^-$ are the sky pixels pointed by the plus-horn and minus-horn
during the observation $k$ respectively.
From  total $L$ observations,  the differential TOD
\boldmath
\begin{equation}
\label{dt1}
d = \mathbf{A}t
\end{equation}
\unboldmath
with $\mathbf{A=}\{a(k,i)\}$ being the scan matrix.
The most of elements $a(k,i)=0$ except for $a(k,i=k^+)=1$
and $a(k,i=k^-)=-1$.
The WMAP team produces the released temperature map \boldmath $\hat{t}$ \unboldmath
from the calibrated TOD  \boldmath $d$ \unboldmath by using
their map-making software~\cite{hin03a,jar07}.

Here we use a simple method to test the consistency between the WMAP TOD
 \boldmath $d$ \unboldmath  and
 the released map  \boldmath $\hat{t}$ \unboldmath reconstructed  from
 \boldmath $d$ \unboldmath by map-making.
We calculate  for each observation $k$ the residual $d'_k$
between the measured calibrated difference and the calibrated difference predicted
by the reconstructed map, $d'_k=d_k-(\hat{t}_{k^+}-\hat{t}_{k^-})$, to get
the residual TOD
\boldmath
\begin{equation}
d'=d-(\hat{t}_+-\hat{t}_-)\,.
\end{equation}
\unboldmath
If the temperature map is properly reconstructed,
only the instrument noise should be left in
\boldmath $d'$ \unboldmath. To check it,
we produce the correlation map of the residual TOD
\boldmath
\begin{equation}
\label{t0} t_0=\mathbf{M}^{-1}\mathbf{A^T} d'\,,
\end{equation} \unboldmath
where  $\mathbf{M=A^TA}$  is diagonally dominant~\cite{hin03a}
\begin{equation}
M^{-1}(i,j)\simeq \frac{1}{N_i}\delta_{ij}
\end{equation}
with $N_i$ being the total number of observations for pixel $i$.
If the temperature map published by the WMAP team is reconstructed correctly,
the correlation  map  \boldmath $t_0$ \unboldmath
should remain only the map-making error
 with low amplitude and no significant structured signal on it.
Because Eq.~\ref{t0} is linear, we have \boldmath
$t_0=\mathbf{M}^{-1}\mathbf{A^T} d - \mathbf{M}^{-1}\mathbf{A^T}
(\hat{t}_+-\hat{t}_-).$ \unboldmath For a correctly reconstructed
map  \boldmath $\hat{t}$ \unboldmath, both \boldmath ${\mathbf
M}^{-1}\mathbf{A^T} d$ \unboldmath and \boldmath $\mathbf
{M}^{-1}\mathbf{A^T} (\hat{t}_+-\hat{t}_-)$ \unboldmath will be
equal to \boldmath $\hat{t}$ \unboldmath, then \boldmath $t_0$
\unboldmath will be exactly zero,
 despite the inevitable numerical computation error, or the map making error.

 \begin{figure}
\begin{center}
\includegraphics[height=5cm, angle=90]{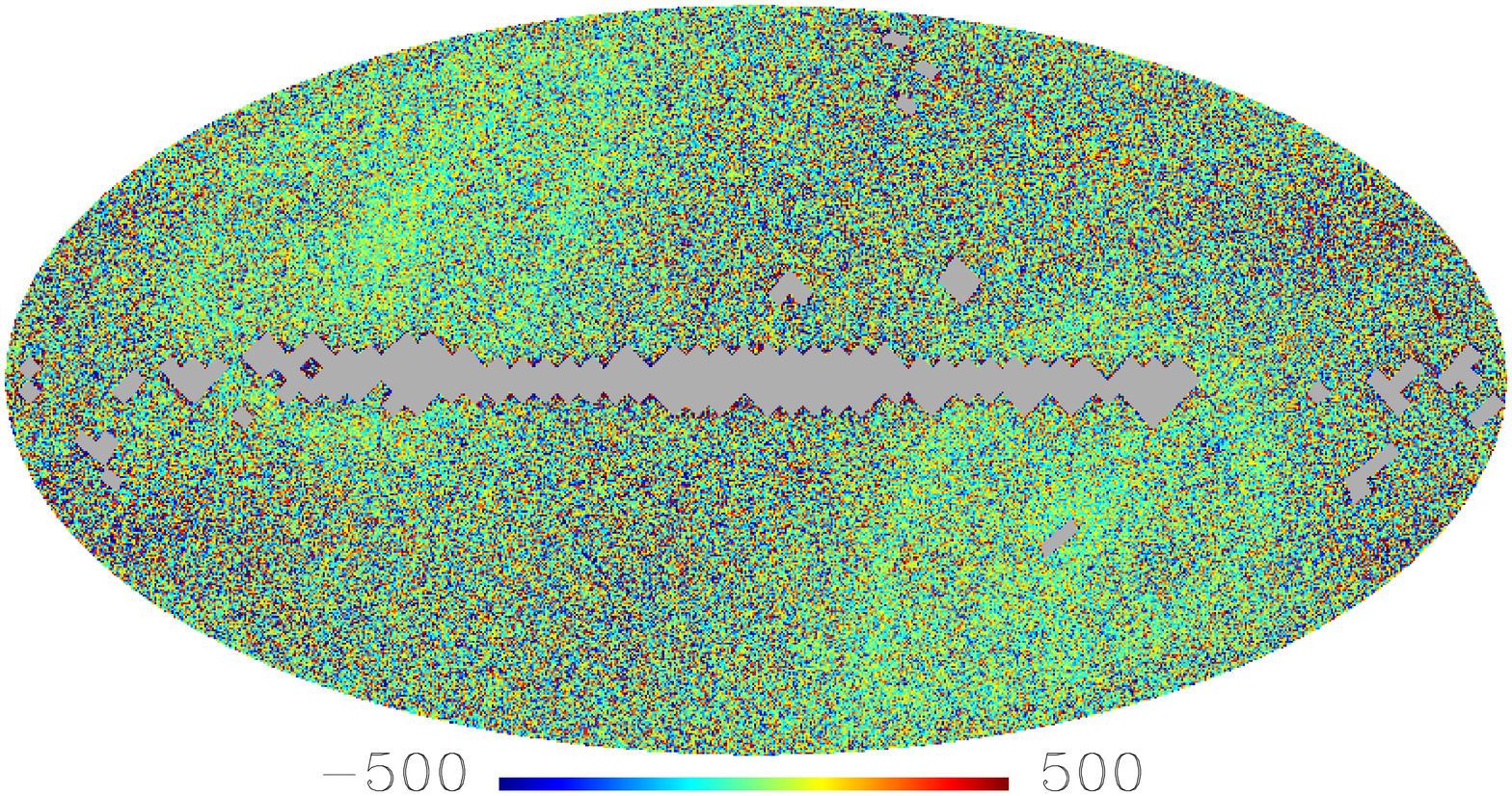}
\includegraphics[height=5cm, angle=90]{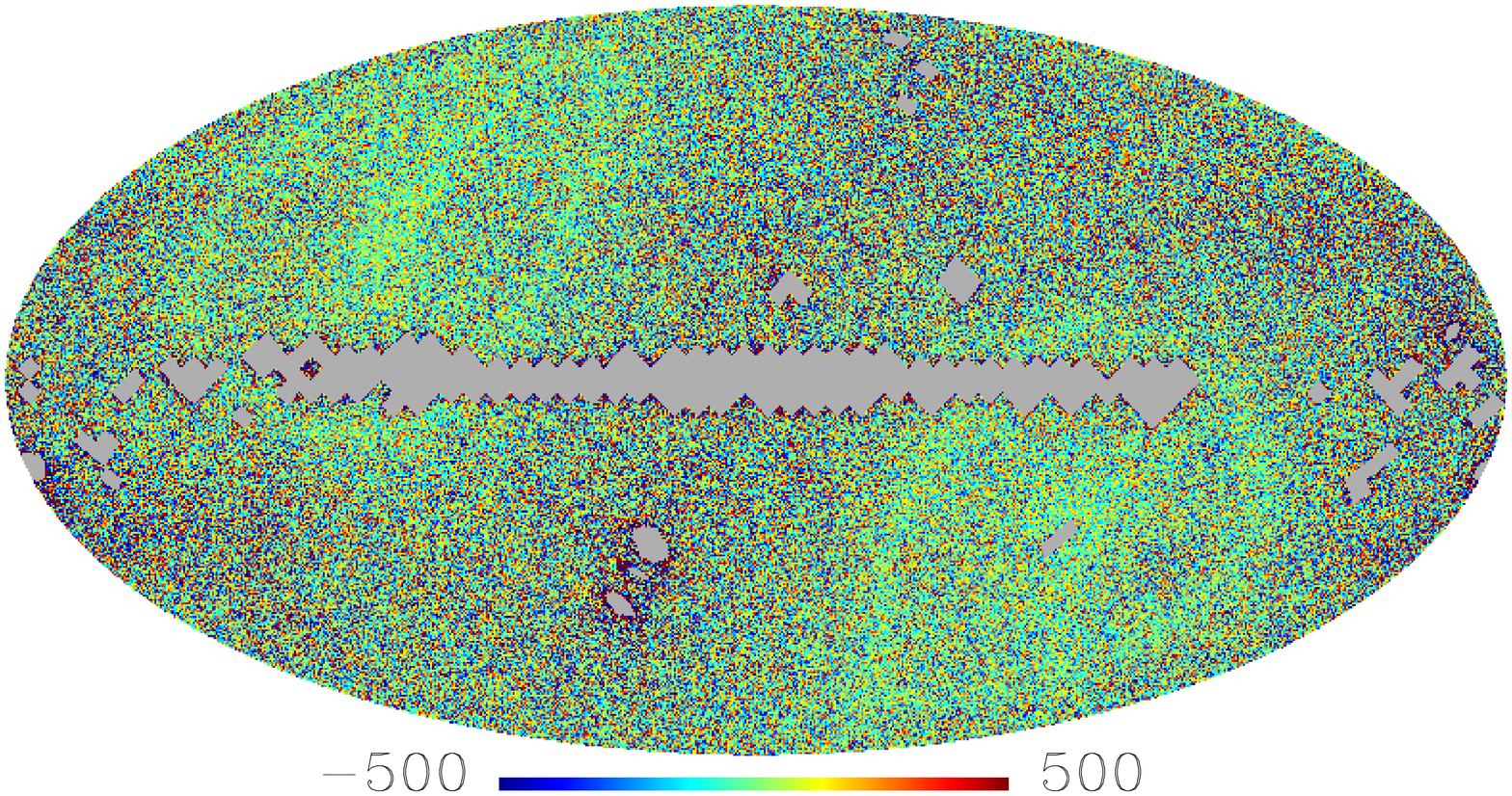}
\end{center}
\caption{  Correlation maps of residual TOD  \boldmath $d'$
\unboldmath between the input TOD and the TOD predicted by the
official WMAP5 year-1 temperature map of Q1-band, in Galactic
coordinates and in units of $\mu \rm{K}$. The gray area represents
pixels without valid value. {\sl Upper panel}: The data used in map
making are those used by the WMAP team to produce the released map.
{\sl Lower panel}: The data used in map making are the measured
calibrated TOD of safe-mode. In this mode, less data are used, see
text for details of this mode.} \label{fig:residual}
\end{figure}

Now we check the consistency of released temperature map with
the used WMAP TOD. In map-making, not all measured TOD are used by the WMAP team.
During the preprocess of map-making, bit-coded
quality flags are set~\cite{lim}. A non-zero flag
indicates that either the observation is problematic in a specific
respect, or the beam boresight is away from one of the out planets
no more than $7\degree$, which is the antennae main beam radius
limit ~\cite{Hill09}. In both cases, the corresponding data are at
least less optimal, and some of these data (not all) are not used
by the WMAP team in the map-making.
We follow the WMAP team's flagging convention\footnote{(1) GENERAL FLAG --
Test on bits 0, 1, 3, 4, 5. This discards data when the observatory
is not in observing mode, the Sun is visible over the shield, but included data
when either the Earth or Moon are visible over the shield (but still in the far
sidelobes of the radiometer beams).
(2) DA-SPECIFIC FLAGS -- Test on bit 0, re-test on bits 1--10.
Bit 0 allows exclusion of data with known thermal disturbances
or radiometer upsets.  Bits 1--10 show that a planet is
close to a radiometer beam.  If bit 1--10 is set,
compute the distance between the indicated planet and the
radiometer beam center based on the instantaneous pointings
for all points within the frame, and discard only those points
for which the planet lies within $1\degree.5$  of the beam.}
to get the used WMAP5 year-1 TOD.
The upper panel of Fig.~\ref{fig:residual} shows the
map \boldmath $t_0$ \unboldmath
obtained by Eq.~\ref{t0} from the used WMAP5 year-1 TOD and
the released WMAP5 year-1 temperatures of Q1-band, where visible structures along the
ecliptic plane and around the ecliptic poles left
 and the rms amplitude $\sim 265\,\rm{\mu K}$,
which is much higher than the expected rms error of $\sim 0.2\,\rm{\mu K}$
estimated by the WMAP team with flight-like simulations
for their map-making algorithm~\cite{hin03a}.

More than $95\%$ of doubtful TOD data are non-zero flagged for the
Sun, Earth, Moon and planet avoidance, the beam boresight angle
for one of the planets is less than
$7\degree$. However, the criterion really
used for the maps released by the WMAP team
is a cut of only $1\degree.5$ ~\cite{lim}, the revealed structure
in the residual TOD
may come from ecliptic contamination by used non-zero flagged TOD.
To test this guess, we use only the high quality data whose quality flags
are all zero (we call this the "safe-mode")
 to produce the residual TOD and its correlation map \boldmath $t_0$. \unboldmath
The high quality TOD (all flags being zero) preserve more than
$74\%$ of all measured WMAP TOD, and the sky coverage of these data
is more than $99.7\%$ (neglecting the WMAP5 processing mask region).
Therefore, these data in safe-mode not only can better avoid the
contamination from the ecliptic plane, but also are enough to
reconstruct  a reliable CMB map as well. This is also ensured by an
end-to-end test upon the total map-making and power-spectrum
estimation processes shown in \S3.4, where they are proved to be
unbiased and highly accurate. The \boldmath $t_0$ \unboldmath map
from the safe-mode TOD, shown in the lower panel of
Fig.~\ref{fig:residual}, still remains a structure similar with what
from the WMAP team used TOD and a large rms amplitude $\sim
291\,\rm{\mu K}$.

The above results of residual TOD test demonstrate
there certainly exists a remarkable problem in the WMAP map-making process
and using carefully tested software to improve the previous
released WMAP results is really needed.

\section{SOFTWARE TESTING}
 We write out our programs for map-making
and power spectrum estimation.
The input of our map-making program is the calibrated
TOD~\cite{lim} downloaded from the WMAP team's website
(fttp://lambda.gsfc.nasa.gov/).
According to the WMAP document~\cite{hin03a}, only three extra
corrections are needed:
 the $1/f$ noise removal, the sidelobe correction,
and the transmission imbalance correction. The sidelobe correction
is applied only to the K-band, and other two corrections are applied
to all bands. To check the reliability of released WMAP results by
reproducing CMB maps from the original raw data independently,
firstly we must test our map making and power spectrum estimation
softwares to make sure that they produce correct outputs. This
includes four tests: the residual TOD test, the residual dipole
component test, the map-making convergence test, and, finally, the
end-to-end test checking all our data pipeline in a self-consistent
manner.

\subsection{\sl Residual TOD}
\begin{figure}
\begin{center}
\includegraphics[height=5cm, angle=90]{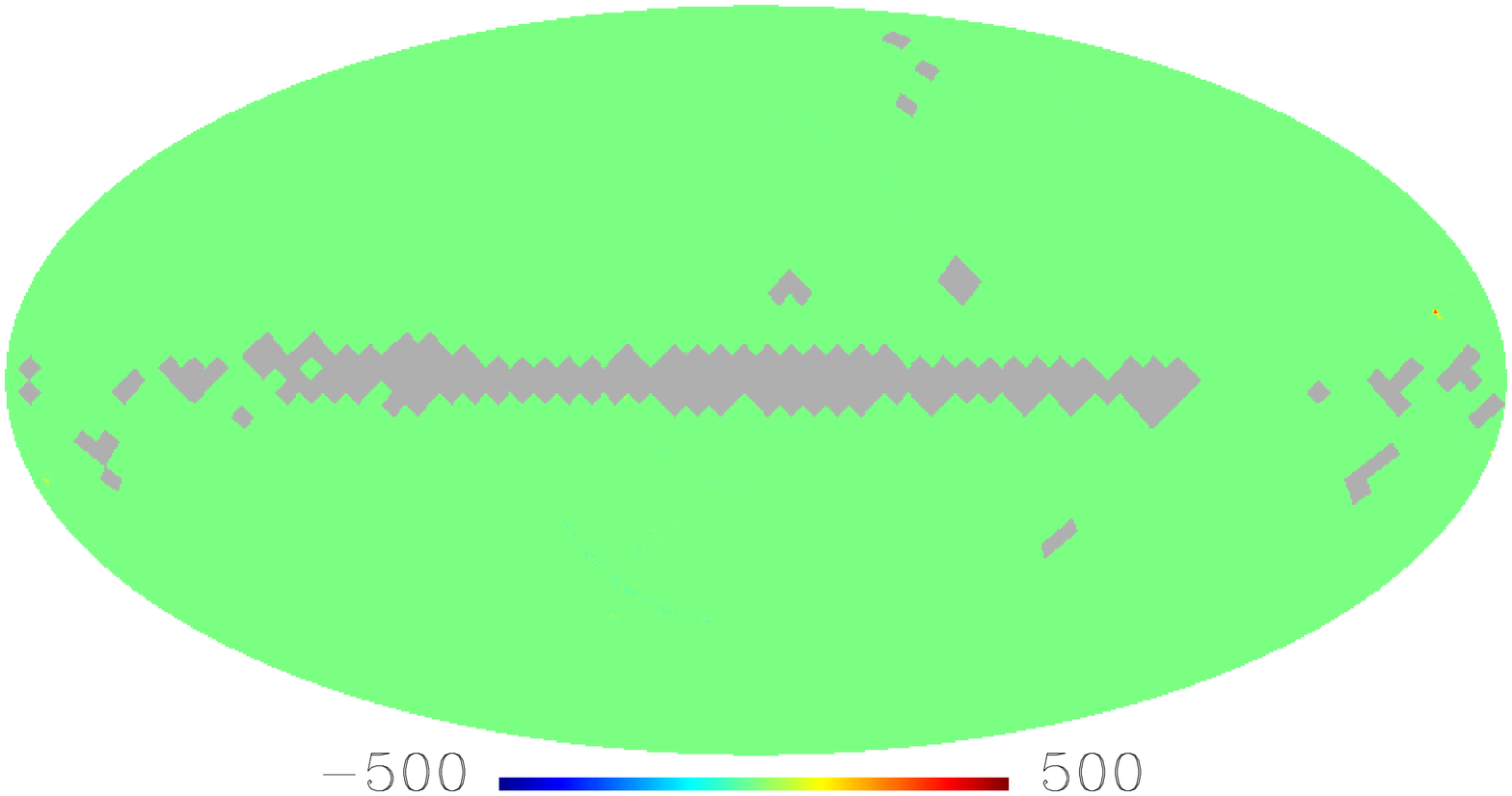}
\includegraphics[height=5cm, angle=90]{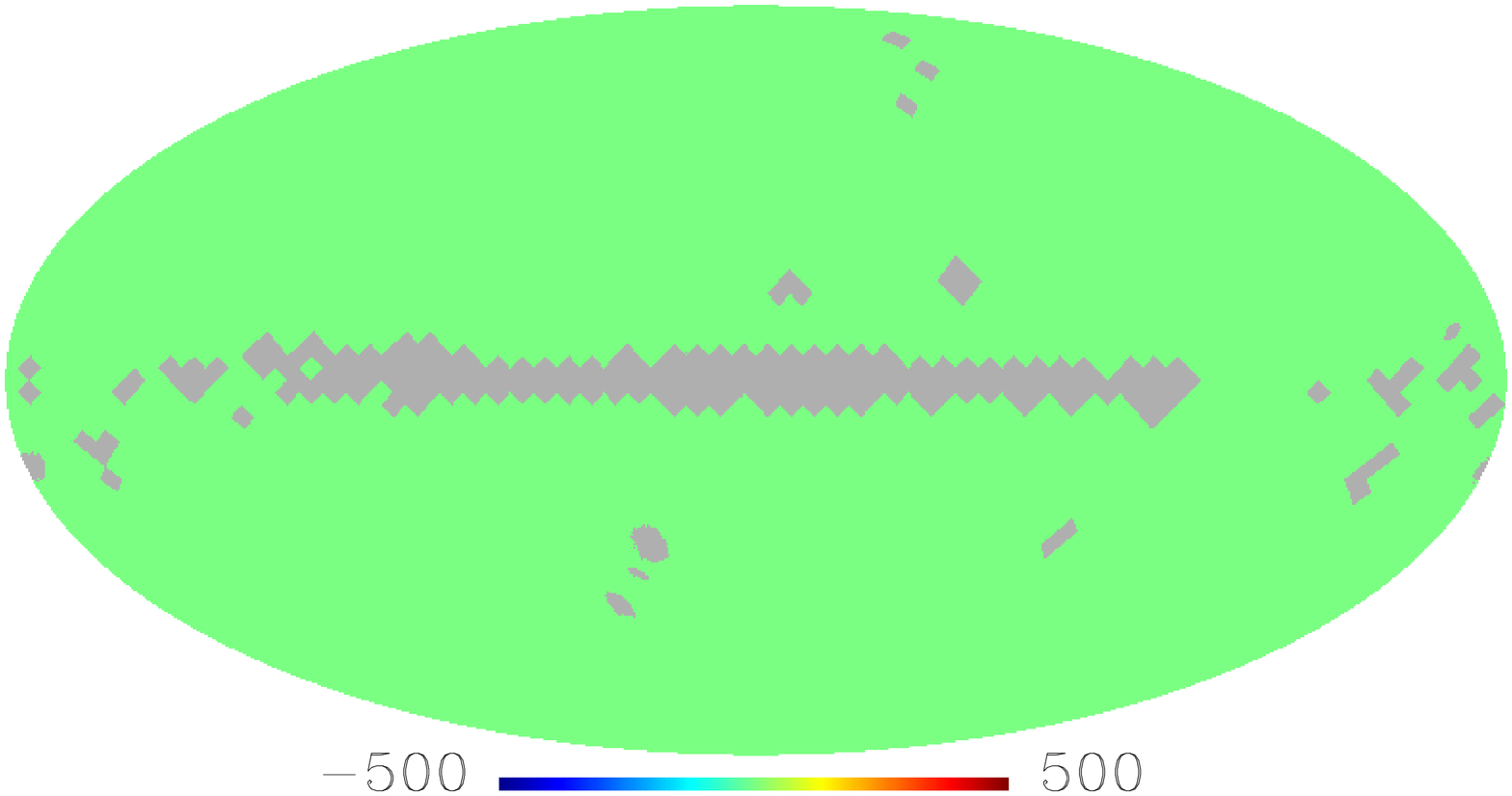}
\end{center}
\caption{  Correlation map of residual TOD between the input TOD and
the TOD predicted by
 our new temperature map of Q1-band, in Galactic coordinates and in units of $\mu \rm{K}$.
The gray area represents pixels without valid value. {\sl Upper
panel}: The data used in map making are those used by the WMAP team
to produce the released map. {\sl Lower panel}: The data used in map
making are the measured calibrated TOD of safe-mode. In this mode,
less data are used, see text for details of this mode. }
\label{fig:residual2}
\end{figure}

With our map-making program we produce new temperature map,
and with the new map and Eq.~\ref{t0} we calculate the residual TOD
as already illustrated. All these are done for both the WMAP5 year-1 Q1-band TOD
and the corresponding safe-mode TOD separately, and their correlation maps are  shown
in the upper and lower panels of Fig.~\ref{fig:residual2} respectively.
In both maps, no significant
structure can be seen and the rms amplitudes are all less than $0.15$ $\rm{\mu K}$,
almost 2000 times lower than what from the WMAP team's map-making products
and close to the rms error of  $\sim 0.11 \rm{\mu K}$
of our map-making (see~\S3.3),
indicating the inconsistent problem presented in \S2 is prevented
in our map-making algorithm.

It has to be noted that we repeat
the map making by writing our own code but following every step given
by the WMAP document~\cite{hin03a}, but do not encounter their problems in our output maps.
This demonstrates that the method of WMAP map-making could in principle
succeed in recovering temperature maps, but in practice
the WMAP team do not implement their own method
correctly.

\subsection{\sl Residual Dipole Component}
The filtered calibrated TOD can not be directly used for map-making,
because they contain large dipole component caused by the joint
motion of the sun and the satellite. This dipole component is
subtracted according to the WMAP team's literature (Hinshaw et al.
2008, \S4). After subtracting the unwanted dipole component, the
residual dipole component amplitude outside the KQ85 mask in our
Q1-band map is estimated to be 7.56 $\rm {\mu K}$ and point to the
Galactic coordinates ($13\degree.16, -11\degree.78 $). Meanwhile,
the residual dipole component amplitude outside the KQ85 mask in the
Q1-band WMAP5 official map is also estimated, which is 7.92 $\rm
{\mu K}$ and points to the Galactic coordinates ($3\degree.93,
-14\degree.70 $). Both these dipole components are estimated with
the "remove\_dipole" program from the standard HEALPix
package~\cite{gor05}, and the difference between them is about 1.1
$\rm {\mu K}$. Since the original dipole component amplitude is
several hundreds times larger than the residual dipole, the
neglectable difference between the two residual dipole components
derived from the new and old CMB maps indicates that both our
map-making and dipole component removal have high accuracy and their
results are reliable.

\subsection{\sl Convergence of Map-making Algorithm}
\label{sub:end-to-end test}

\begin{figure}
\begin{center}
\includegraphics[height=5cm, angle=90]{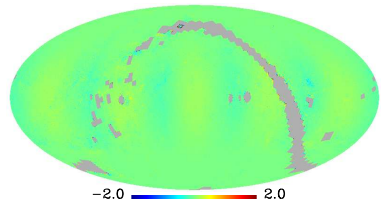}
\end{center}
\vspace{-2mm} \caption{  Map-making error, $t_{out}-t_{in}$
($\rm{\mu K}$), in ecliptic coordinates by our map-making software
after 50 rounds of iterations. The input map is Q1-band WMAP5
temperature map without foreground removal, and the map-making is
done with our software in safe-mode. All observations with either
beam inside the WMAP5 processing mask are rejected in addition to
the safe-mode data rejection. The rms of $t_{out}-t_{in}$ in this
map is $\sim 0.11 \rm{\mu K}$. The gray area represents pixels
without valid value.} \label{tod_residual}
\end{figure}

The map-making algorithm is an iterative one, thus it is
important to test its convergence. An official test had been adopted
by the WMAP team~\cite{hin03a}. They generated artificial TOD from
simulated CMB map ($t_{in}$) according to the scan scheme of the
real observations, then use the artificial TOD to do map-making and
compare the result $t_{out}$ with $t_{in}$ to see if their software
converges well. The output $t_{out}$ is expected to be very close to
the known input $t_{in}$, and the rms of $t_{out}-t_{in}$ is
expected to decrease as the number of iterations increases, right as
presented in the WMAP team's literature (Hinshaw et al. 2003a, Fig.
2, 50 rounds of iterations).

We repeat this test with our map-making software for four
individual tests: (1) $t_{in}$ being real WMAP temperature maps and
the scan scheme being safe-mode; (2) similar to the above test but
the scan scheme being all observations; (3) and (4) are similar to
(1) and (2) respectively, but $t_{in}$ being simulated CMB maps plus
WMAP-like foreground. Each test was run for 50 rounds of iterations.
Theoretically speaking, the results of these four tests should be
very close to each other, because the inputs and map-making
environments do not differ much among them. The test results are
right as expected: The maps of $t_{out}-t_{in}$ given by all the
four tests are similar to the WMAP team's official test (Hinshaw et
al. 2003a, Fig.~2), and all the four rms of $t_{out}-t_{in}$ are
less than 0.2\,$\rm{\mu K}$, which do not exceed the residual rms
given by the WMAP team. The result of test (1) for Q1-band is
presented in Fig.~\ref{tod_residual}. The results of tests (2)--(4)
are similar to Fig.~\ref{tod_residual}. With such tests, we are
confident that our software is able to make reliable CMB maps from
the WMAP TOD.

\subsection{\sl End-to-End Test}

\begin{figure}
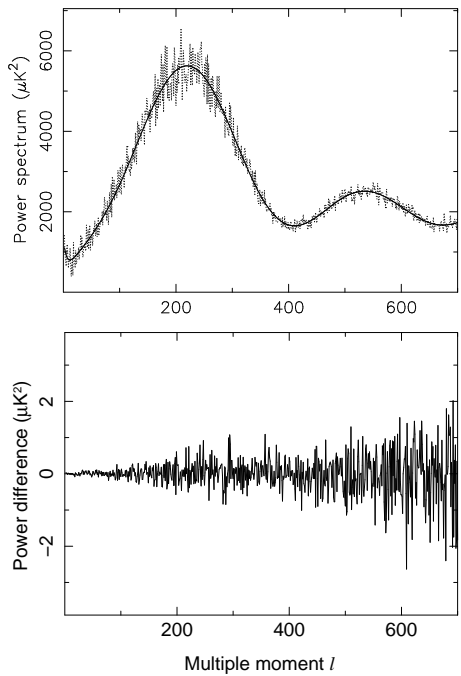

\begin{center}
\includegraphics[height=6.cm, angle=270]{f4a.ps}\\
\vspace{2mm}
\includegraphics[height=6.cm, angle=270]{f4b.ps}
\end{center}
\caption{ End-to-end test. {\sl Upper panel}: the input $\rm{\Lambda
CDM}$ power spectrum (solid line) and the final cross power spectrum
(dotted line). {\sl Lower panel}: power differences between the
final cross-power spectrum and the cross power spectrum generated by
synfast, $C_{final}-C_{synfast}$.} \label{fig:end-to-end test}
\end{figure}

This is the most important test, in which the overall map-making and
power spectrum estimation are checked in an end-to-end way: starting
from the best-fit WMAP5 $\rm{\Lambda CDM}$ power spectrum
($C_{\Lambda CDM}$), we use the "synfast" routine in HEALPix software package
(available at http://healpix.jpl.nasa.gov) to generate
simulated sky maps for V and W-bands, and then use these maps to
produce artificial TOD in safe-mode. These artificial TOD are used
to do map-making and then the recovered maps by the map-making
process are used to compute the final cross-power spectrum
($C_{final}$). In the end, $C_{final}$ is compared to $C_{\Lambda
CDM}$ to finish the end-to-end test.

Such a test can definitely illustrate two things: whether the $\sim
74\%$ observations used in map-making are enough to
recover an unbiased CMB map, and whether our data pipeline is
unbiased and accurate. If the final power spectrum is consistent
with the input $\rm{\Lambda CDM}$ power spectrum, then we can say
that our entire data processing has been tested for all aspects and
been well justified.

We give the results of this test in Fig.~\ref{fig:end-to-end test}.
More than $99\%$ of $C_{final}-C_{\Lambda CDM}$ is attributed to the
random fluctuation introduced automatically by synfast to simulate
the cosmic variance. To isolate and show the tiny error attributed
to the overall map-making and power spectrum estimation processes,
we derive cross power spectrum from the V and W-bands maps generated
by synfast ($C_{synfast}$) with the same power spectrum estimation
program and compare $C_{final}$ with $C_{synfast}$ in the lower
panel of Fig.~\ref{fig:end-to-end test}. This is the true variation
introduced by the overall map-making and power spectrum estimation
processes.

For $l=2$\,--\,700, the average
is only $0.0052$ $\rm{\mu k^2}$ and rms only $0.51$ $\rm{\mu k^2}$
for $C_{final}-C_{synfast}$,
the  average is only $-6.49$ $\rm{\mu
k^2}$ and rms  is $234.96$ $\rm{\mu k^2}$ for $C_{final}-C_{\Lambda CDM}$.
Therefore, we see no systematical bias in either
$C_{final}-C_{synfast}$ or $C_{final}-C_{\Lambda CDM}$, and the
overall map-making and power spectrum estimation has high accuracy.
In other words, our map-making and power spectrum estimation has
passed the end-to-end test and has been well justified.

\begin{figure}
\begin{center}
\includegraphics[height=5cm, angle=90]{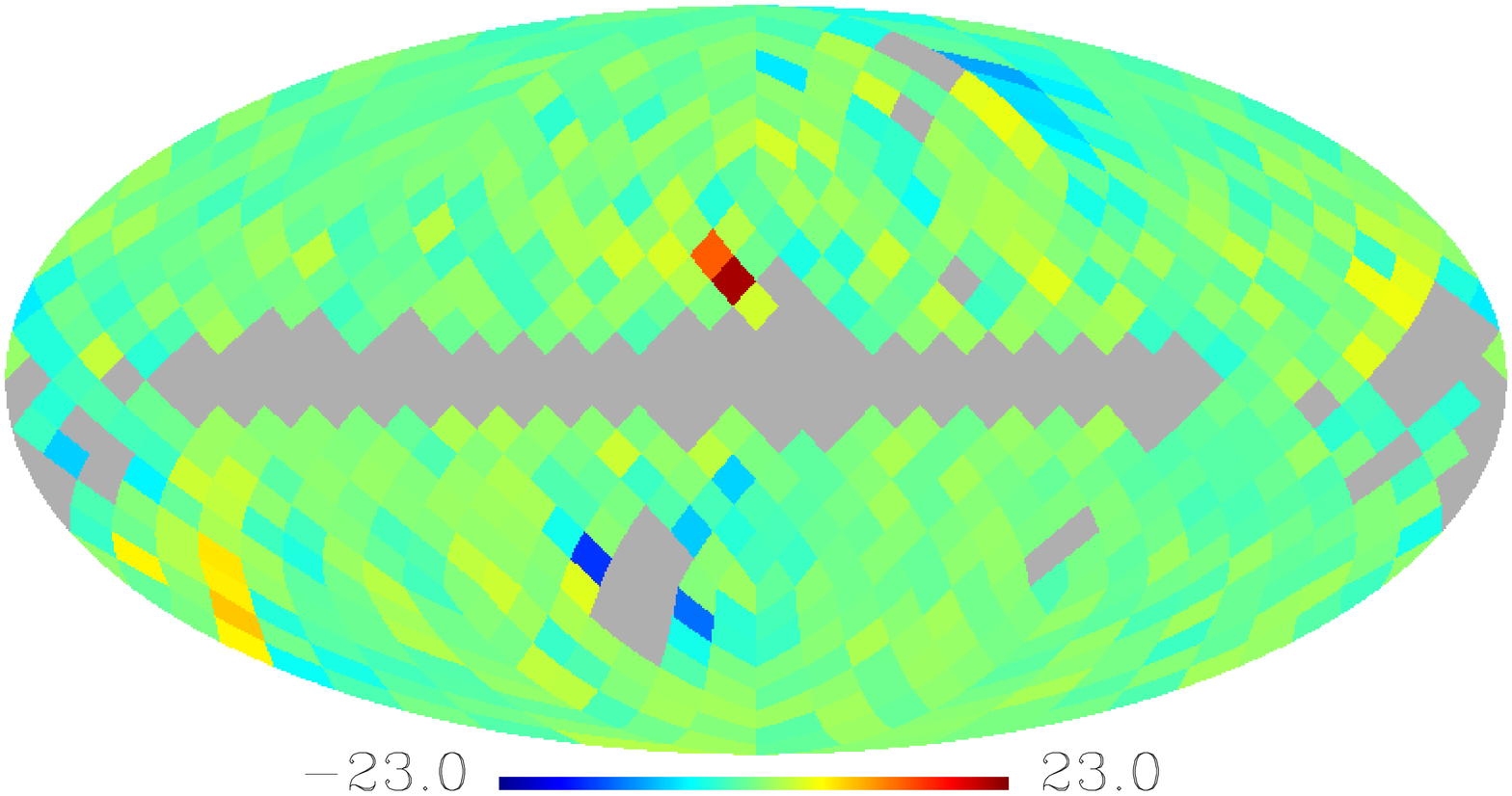}
\includegraphics[height=5cm, angle=90]{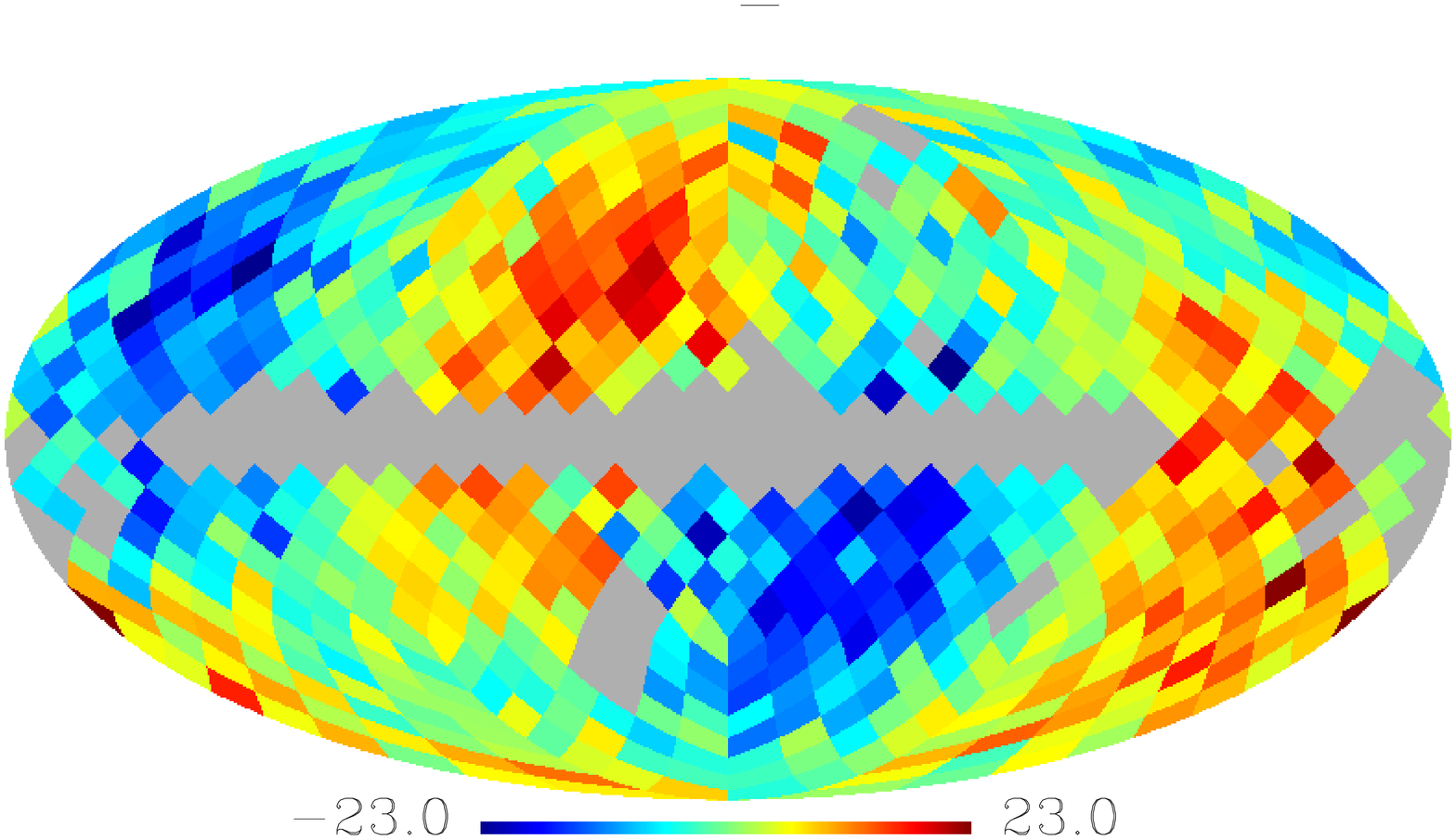}
\end{center}
 \caption{ Downgraded (to $N_{side}=8$) Q1-band
difference maps in Galactic coordinates and in units of $\mu
\rm{K}$. The gray area represents pixels without valid value. {\sl
Upper panel}: the map produced with our map-making software from the
TOD used to make the published WMAP5 year-1 map minus the
corresponding safe-mode map. {\sl Lower panel}: The WMAP5 year-1
official map minus the corresponding safe-mode map.}
\label{fig:dif-map}
\end{figure}

\begin{figure*}
\begin{center}
\includegraphics[height=5.cm, angle=90]{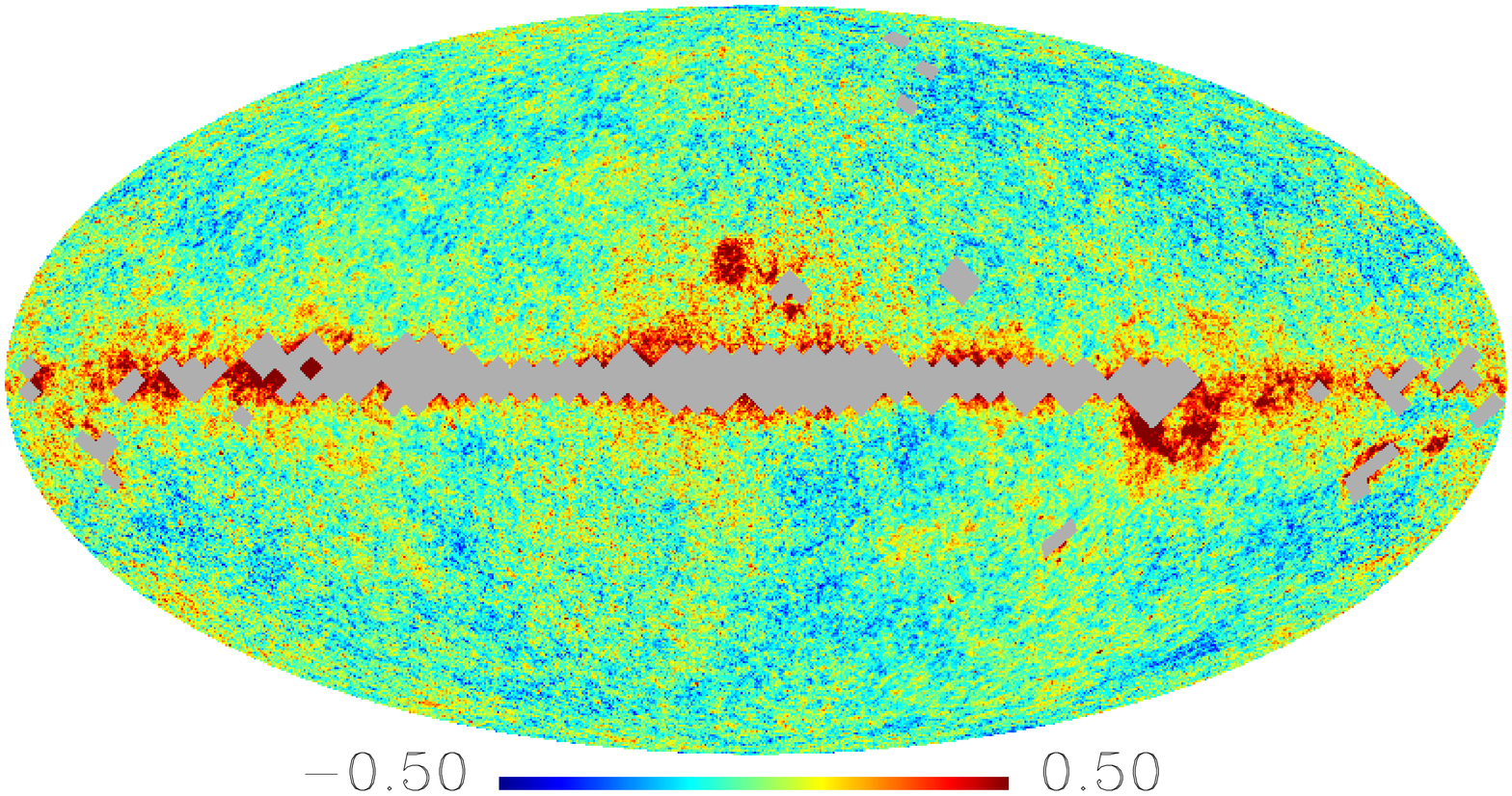}
\hspace{1mm}\includegraphics[height=5.cm, angle=90]{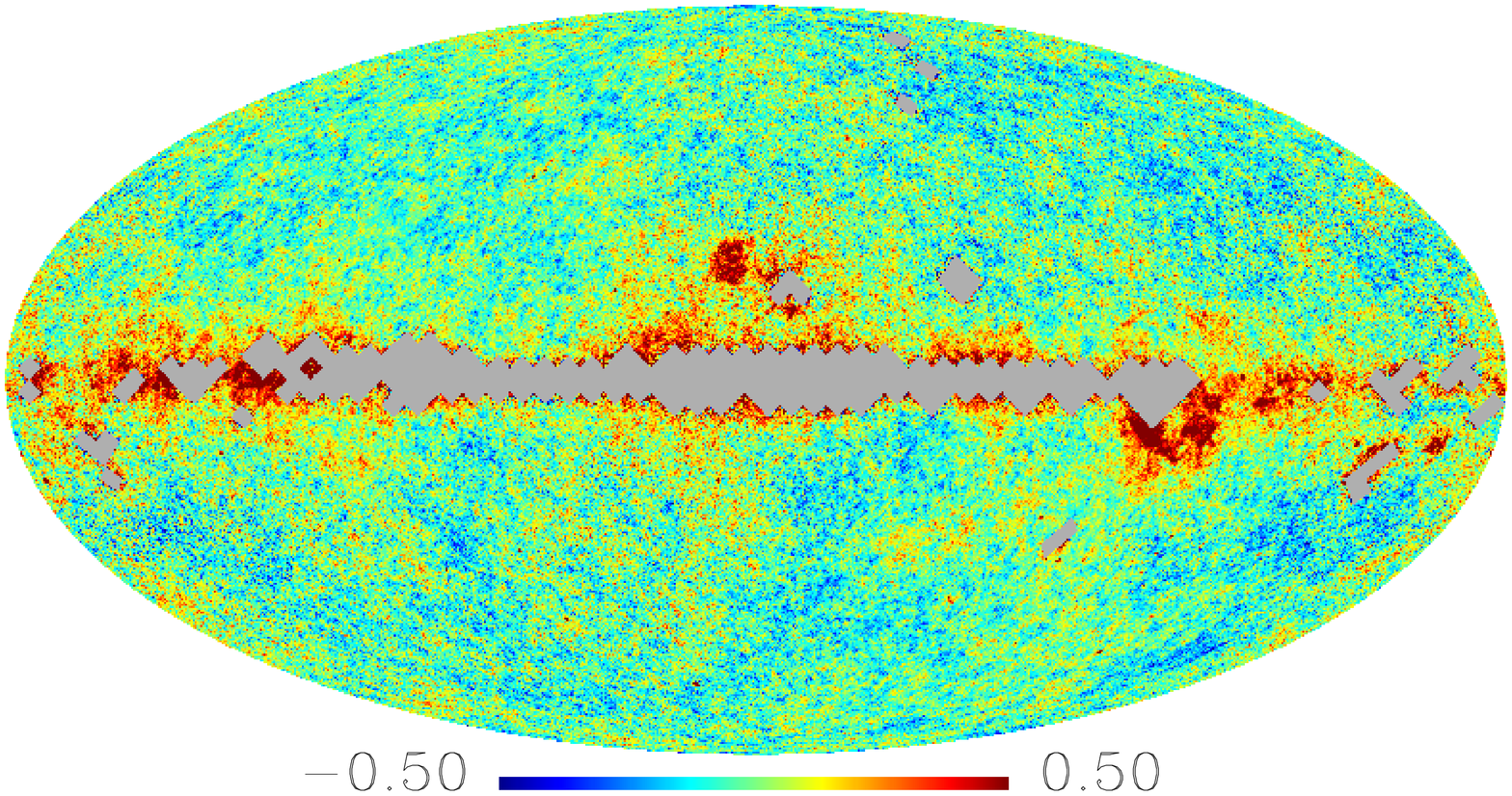}
\hspace{1mm}\includegraphics[height=5.cm, angle=90]{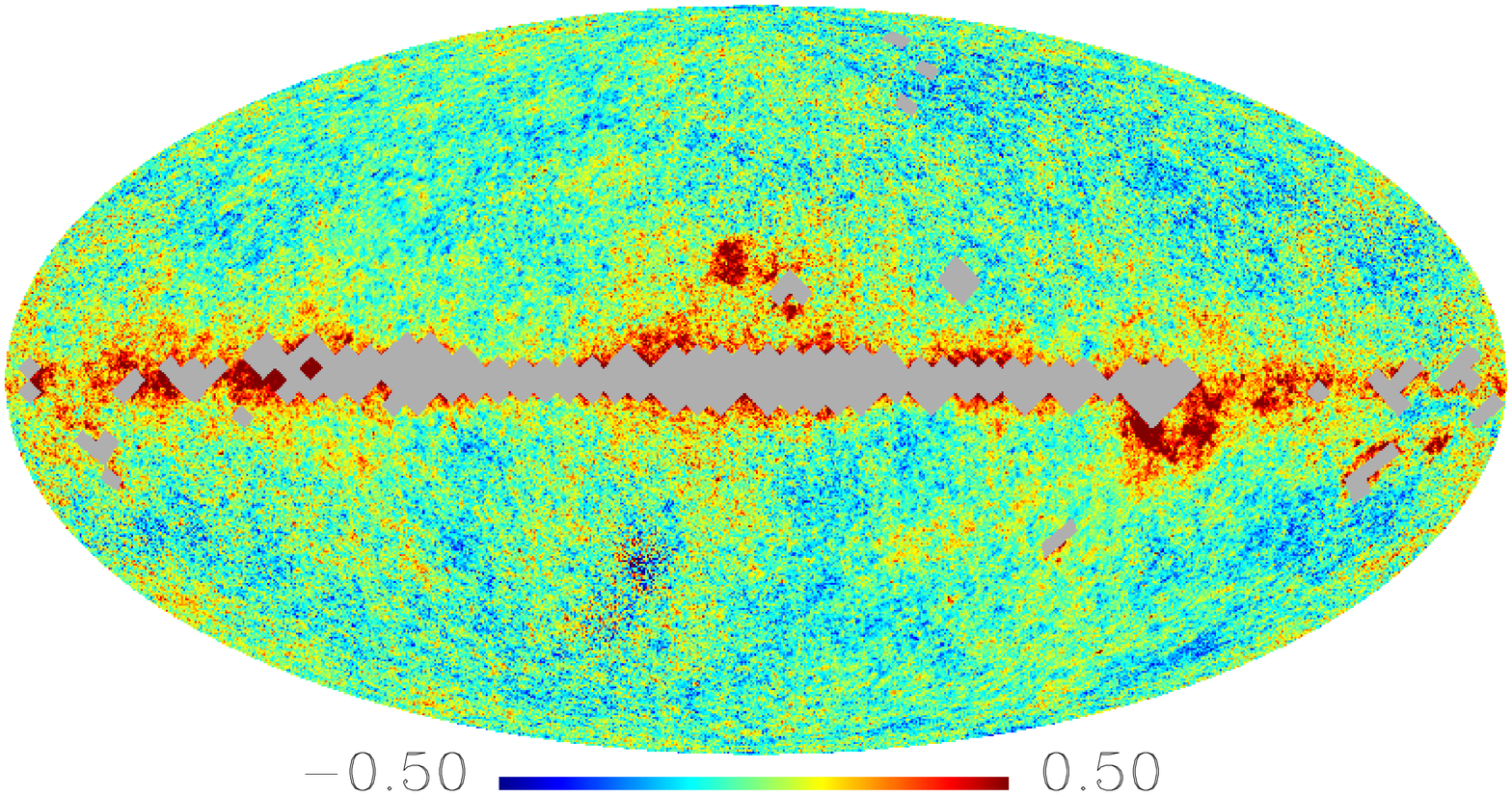}
\includegraphics[height=5.cm, angle=90]{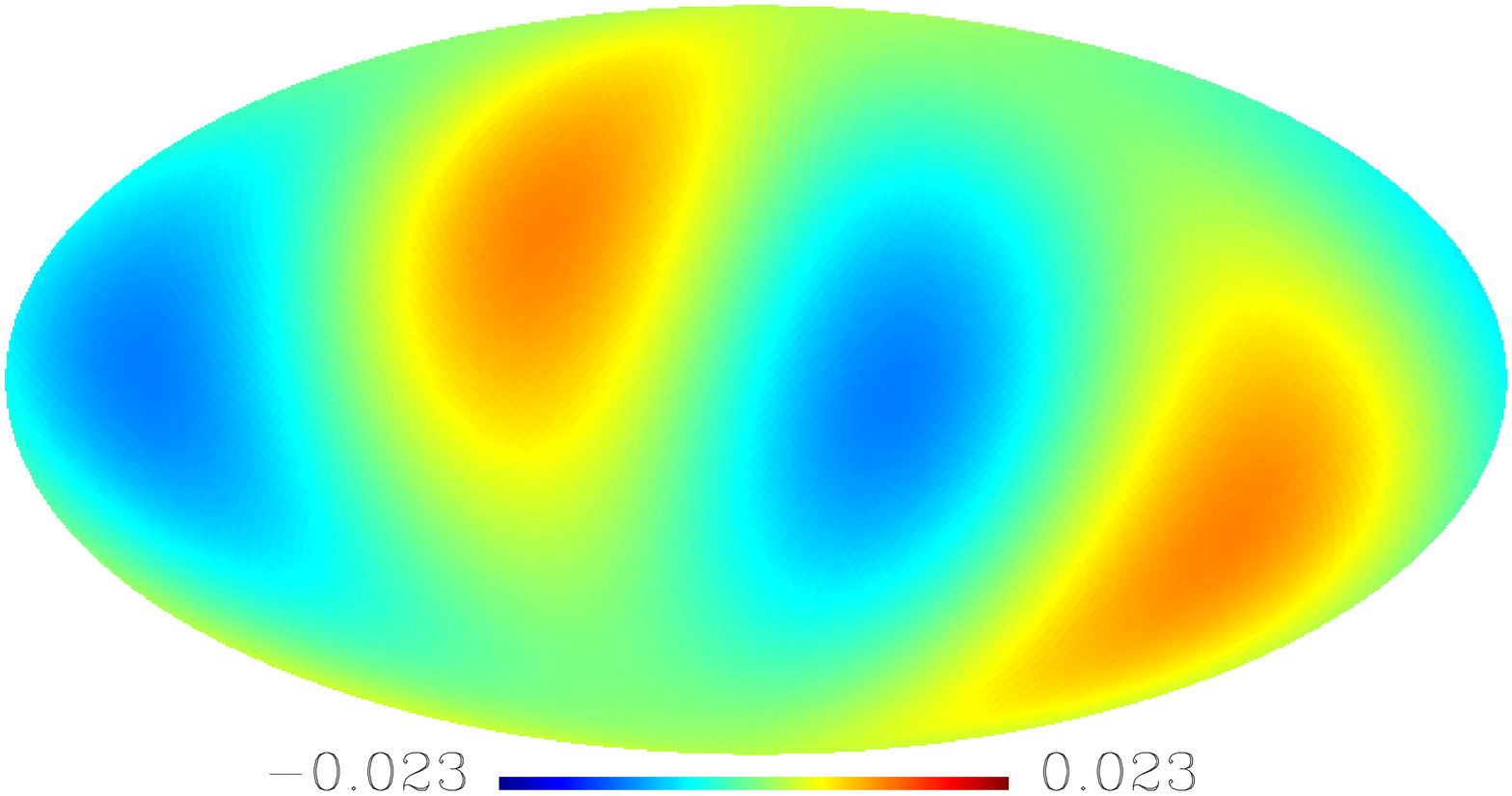}
\hspace{1mm}\includegraphics[height=5.cm, angle=90]{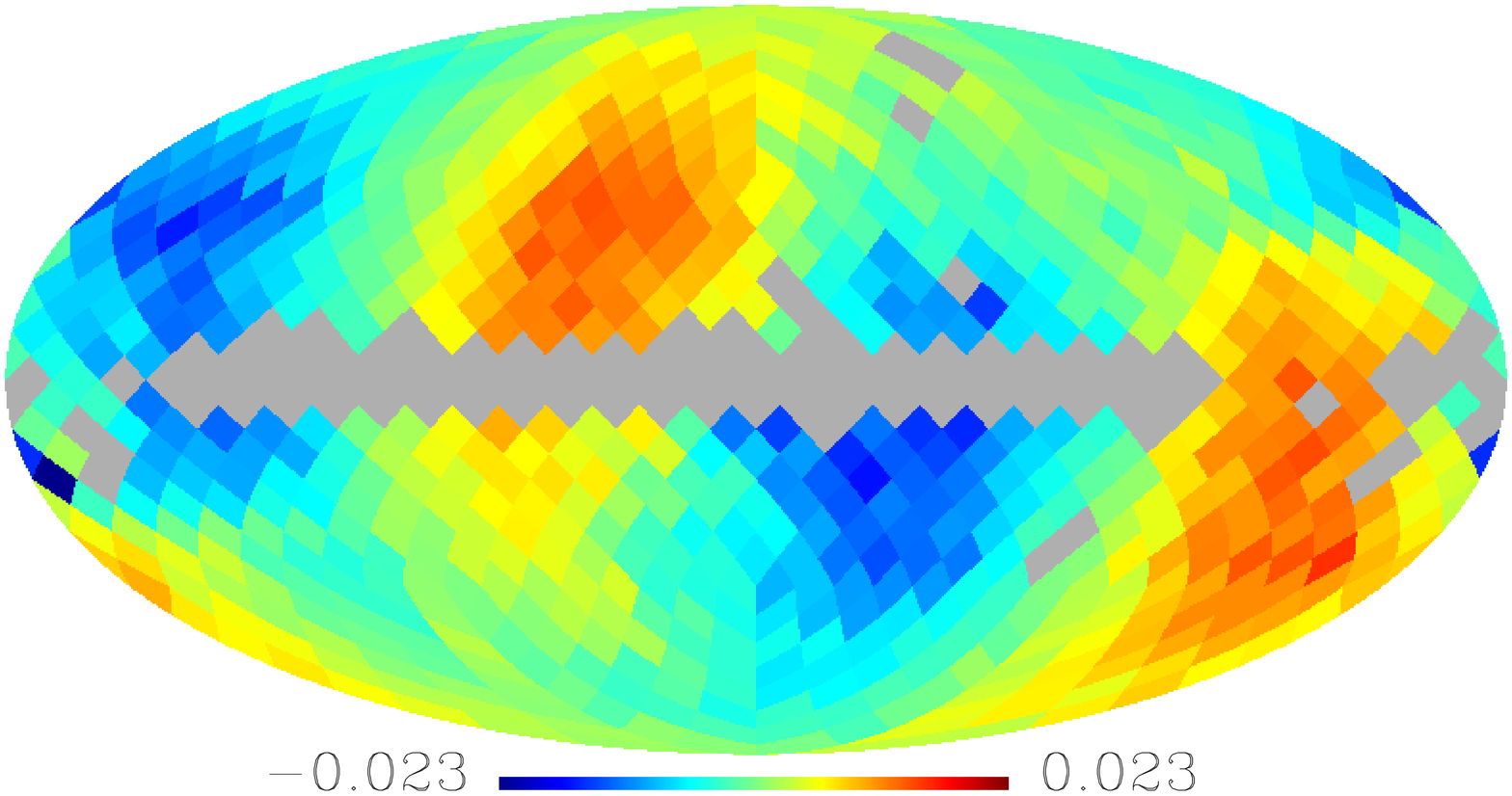}
\hspace{1mm}\includegraphics[height=5.cm, angle=90]{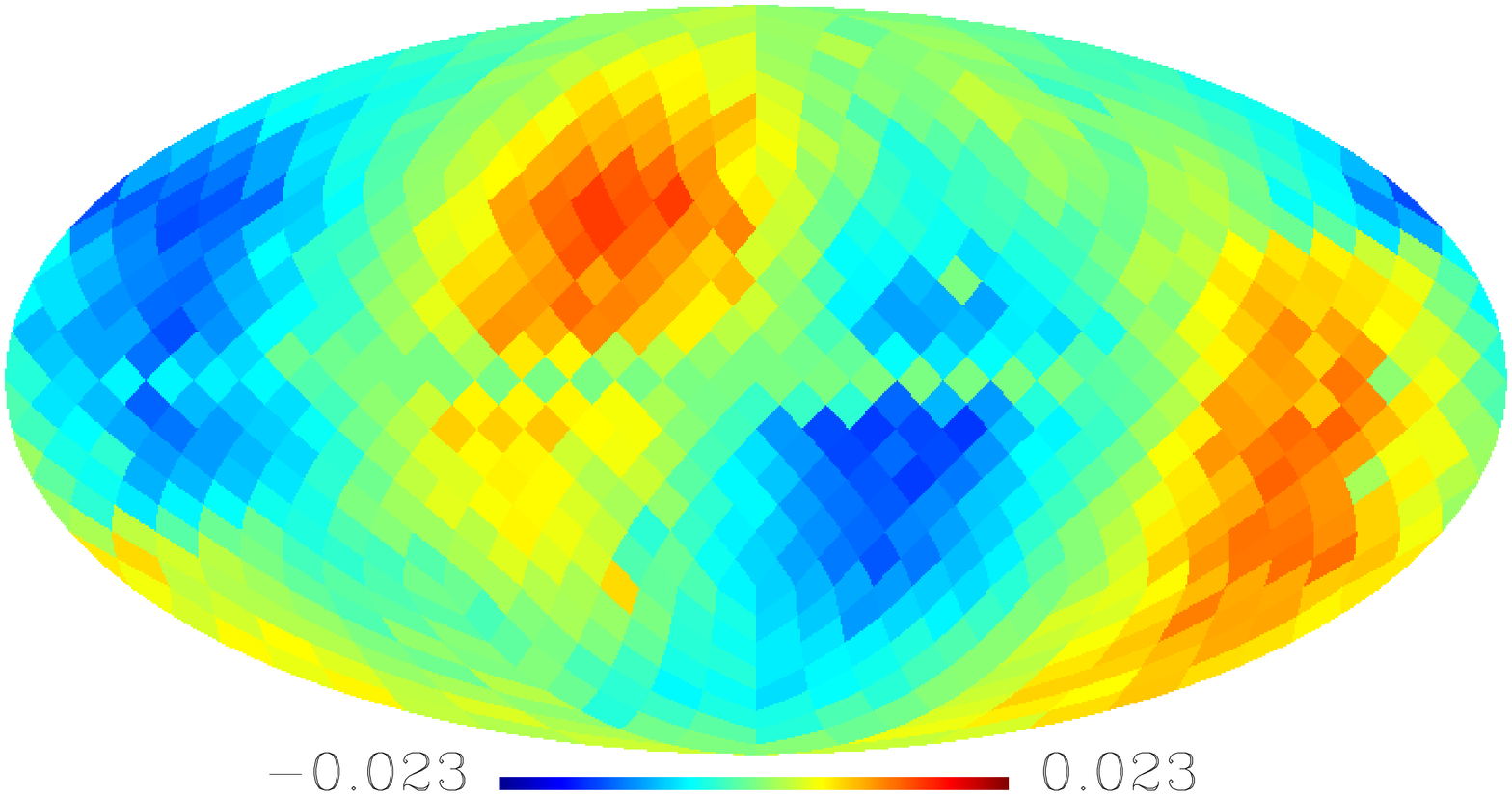}
\end{center}
\vspace{-1mm} \caption{ {\bf Upper panel} -- Q1-band temperature
maps in units of mK and in Galactic coordinates.
 The monopole and dipole components have been removed
for all maps. {\sl From left to right}: the official WMAP5 map; the
new WMAP5 map reconstructed from the TOD used in WMAP5 map-making
but by our software; the safe-mode map reconstructed from high
quality WMAP5 TOD by our software. {\bf Lower panel} -- {\sl Left}:
the quadrupole component of the official WMAP5 CMB map; {\sl
Middle}: the difference between the official WMAP5 map and the new
WMAP5 map (smoothed to $N_{side}=8$); {\sl Right}: the difference
between the official WMAP5 CMB map and the safe-mode map
 (smoothed to $N_{side}=8$).
} \label{fig:map}
\end{figure*}

The correlation maps of residual TOD shown in \S3.1 demonstrate that our
map-making product is consistent with the input TOD, but the published WMAP maps
contain remarkable constructed error component produced during map-making.
The results of testing residual dipole component amplitude
(\S3.2) and convergence of iterative map-making algorithm (\S3.3) show that
our software has good performance.
The end-to-end simulations (\S3.4) give a prove that all our data pipeline
is comprehensively self-consistent.
We therefore expect the temperature map recovered from WMAP TOD by our software
can provide more reliable result for CMB study.

It is worth to note that, with our map-making program, we also
reconstruct temperature maps from the TOD used to produce the
released WMAP5 map. We find that, despite a few minor differences,
the obtained map is much more consistent with our safe-mode map than
with the WMAP team's official map, as presented in
Fig.~\ref{fig:dif-map}, which illustrates the safe-mode being not
the reason of the major difference between the WMAP release and our
result.

\section{TEMPERATURE MAPS}
\label{sub:the safe mode maps and old maps}
With our map-making software we reconstruct sky temperature maps
from WMAP TOD.
In this work we discuss and solve for the temperature only, which is
the average of all four WMAP channels: $d=(d_{13}+d_{14}+d_{23}+d_{24})/4$.
The final temperature maps are derived from TOD with 80 rounds of
iterations. The map-making is done in $N_{side}=512$ resolution~\cite{gor05} and
the HEALPix nested pixelization scheme only. Observations that have
either beam inside the WMAP5 processing mask are rejected in our
map-making, hence all pixels inside the WMAP5 processing mask have
no values in the final temperature map. However, this does not
affect the power spectrum estimation, because the WMAP5 processing
mask is completely covered by the mask used for power spectrum
estimation (the KQ85 mask or the enlarged KQ85 mask, which is
discussed in \S\ref{sec:ABOUT THE POWER SPECTRUM ESTIMATION}), the
pixels with no values will not be introduced into the power spectrum
estimation.

The upper-left plot of Fig.~\ref{fig:map} is the WMAP5 official
release for the Q1-band, and the upper-middle plot shows the new CMB
temperature map of the Q1-band derived from the WMAP5 used TOD
by our map-making software (both have the best-fit dipole and monopole
components removed, and the plotted temperature ranges are -0.5 --
0.5 mK). The differences between them (official WMAP5 map minus our map) are
computed. We
downgrade the difference map to a low resolution map with the
resolution parameter $N_{side}=8$ and show in the lower-middle plot
of Fig.~\ref{fig:map}.

The rms temperature fluctuation on the smoothed difference map in
the lower-middle plot of Fig.~\ref{fig:map} is $6.6$ $\rm{\mu K}$,
which is much higher than all known map making errors.
We notice that, the downgraded difference map in the lower-middle
plot of Fig.~\ref{fig:map} is almost the same with the WMAP5 quadruple
component (lower-left of Fig.~\ref{fig:map}), which indicates
the quadruple component in the WMAP5 CMB map is almost completely artificial.

We also reconstruct a safe-mode map from the WMAP5 Q1-band TOD of safe-mode
by our map-making software and compute its difference to the
official WMAP5 map, the results, shown in the upper-right
and lower-right plots of Fig.~\ref{fig:map} respectively,
are very similar to what
produced from the WMAP5 used TOD, demonstrate again the observed remarkable
difference is mainly produced by the WMAP map-making process.

\section{CMB ANGULAR POWER SPECTRUM}

\subsection{\sl Estimation Method}
\label{sec:ABOUT THE POWER SPECTRUM ESTIMATION}
\begin{figure}
\begin{center}
\includegraphics[height=5cm, angle=90]{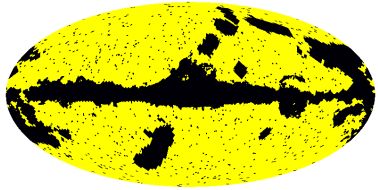}
\end{center}
\vspace{-0.4cm} \caption{  The enlarged KQ85 mask used in computing
single year cross-power spectra. }\label{mask}
\end{figure}

To produce a reliable CMB power spectrum, the foreground emission
should first be subtracted from a raw temperature map to make the clean
CMB temperature map.
Our single-year clean maps are produced with the foreground removal
technique used by the WMAP team~\cite{hin07,gol09}, then the best-fit dipole and
monopole components are subtracted before power spectrum estimation.
After these operations, 30 V and W-bands single-year
temperature maps are used to compute 435 cross power spectra
with the KQ85 mask (same
as WMAP team's five-year cross power spectra estimation scheme).
For the safe-mode maps, due
to data rejection pixels in some areas of the
single-year maps contain no usable observation.
To exclude these pixels (as well as
pixels with too few observations) in computing the CMB power
spectra, an extra mask that mainly lies on the ecliptic plane is
used in addition to the KQ85 mask (which is the default mask for
WMAP team's power spectrum estimation) to constitute an enlarged
KQ85 mask as shown in Fig.~\ref{mask}.

The released WMAP CMB power spectrum is derived from temperature
maps via several main steps including the cross-power spectrum
estimation, cut-sky correction, beam function correction, and pixel
window correction~\cite{hin03b}. Two software packages, HEALPix and
PolSpice, are designed to do these~\cite{gor05,Sz01,chon04}. The
main difference between them is that HEALPix does not support the
cut-sky correction directly, but PolSpice does. Therefore, we use
PolSpice in this work to produce power spectra from our
single-year maps. HEALPix are also used with the cut-sky corrections
done by our own software, and the results are almost identical with
PolSpice results.

\subsection{\sl Power Spectrum}
\label{sub:ps and uncertainty}
\begin{figure}
\begin{center}
\includegraphics[height=7.5cm,angle=270]{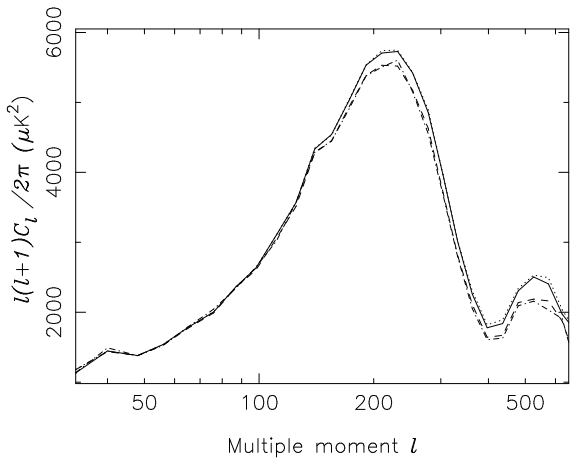}
\end{center}
\vspace{-2mm} \caption{ Binned cross-power spectra ($l=33$ -- 675).
{\sl Solid line}: the released WMAP5 spectrum. {\sl Dotted line}:
the spectrum derived by us from released WMAP5 maps to test our
power spectrum estimation processes (very close to the solid line
and is hard to distinguish). {\sl Dashed line}: the spectrum of new
CMB maps from WMAP5 used TOD and our map-making software. {\sl
Dashed-dotted line}: the spectrum of safe-mode CMB maps from WMAP5
safe-mode TOD and our map-making software. }\label{power spectrum}
\end{figure}

For a start, to test our power spectrum estimation processes,
we derive a binned power spectrum from the released WMAP5 CMB maps
with the same binning scheme as the WMAP team used.
The obtained spectrum is nearly identical with what provided
by the WMAP team (see the solid line and dotted line
in Fig.~\ref{power spectrum}, where the
two WMAP5 power spectra obtained by the WMAP team and by us
respectively are so close to each other that it is hard to distinguish them).
Then we compute the binned cross power spectrum from our new CMB maps
with WMAP5 used TOD and that with WMAP5 safe-mode TOD respectively,
the results
are shown  in Fig.~\ref{power spectrum} for high order moments ($l=33$ -- 675)
and Fig.~\ref{lowl} for low order moments ($l<30$) respectively.

\begin{figure}
\begin{center}
\includegraphics[height=7.5cm,angle=270]{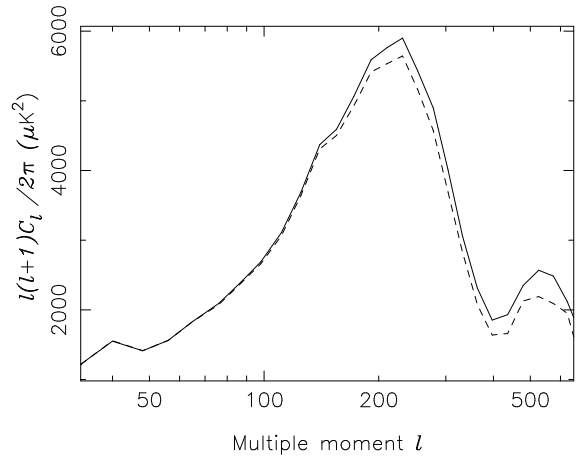}
\end{center}
\vspace{-2mm} \caption{ Binned cross-power spectra from raw maps
before foreground subtraction. {\sl Solid line}: from the released
WMAP5 maps. {\sl Dashed line}: from our safe-mode maps.
}\label{dirty-spectrum}
\end{figure}

We also compute binned cross-power spectra from raw maps
 before foreground subtraction for WMAP5 release and for our safe-mode maps separately,
the results are shown in Fig.~\ref{dirty-spectrum}. From
Fig.~\ref{dirty-spectrum} we can see that the remarkable difference
between the official and our new spectra does not come from the
foreground removal process.

\subsection{\sl Low Order Moments}
\begin{figure}
\begin{center}
\includegraphics[height=6.cm, width=5.5cm, angle=270]{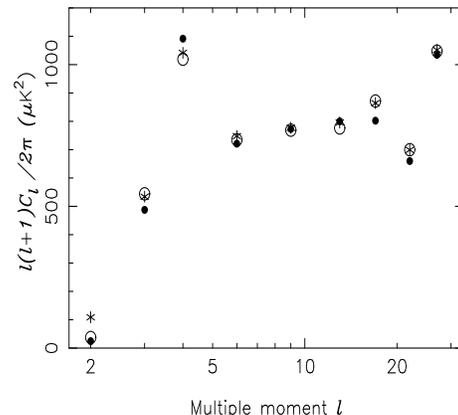}
\end{center}
\vspace{-4mm} \caption{ Low-$l$ cross-power spectra ($l<30$). {\sl
Asterisk}: the released WMAP5 spectrum. {\sl Circle}: the spectrum
of new CMB maps  from WMAP5 used TOD and our map-making software.
{\sl Filled circle}: the spectrum of safe-mode CMB maps from WMAP5
safe-mode TOD and our map-making software.
 }\label{lowl}
\end{figure}

Fig.~\ref{lowl} shows large-scale binned CMB power
spectra for multiple moment $l< 30$, where asterisks mark
WMAP5 release, circles and filled circles mask our results
from new WMAP5 maps and safe-mode maps respectively.

The remarkable quadruple structure in the difference map reminds us
to see whether the large scale anomalies detected in the released
WMAP CMB power spectrum are really cosmic origin or just an
observation effect for some artificial reasons. A lack of anisotropy
power on the largest angular scales has long been observed in WMAP
CMB maps~\cite{ben03b,hin03b,hin08}. We see from Fig.~\ref{lowl}
that the power at $l=2$ further drops to nearly zero in our CMB maps
($37.3\,\rm{\mu K}^2$ from our new WMAP5  maps, $24.4\,\rm{\mu K}^2$
from our safe-mode maps, vs. $108.7$ $\rm{\mu K}^2$ from WMAP5
release)\footnote{The WMAP5 power spectrum and our power spectrum
here are the pseudo-$C_l$ based cross power spectrum.}; meanwhile,
the power spectrum uncertainty attributed to measurement errors at
$l=2$ is $\sim 5.60$ $\rm{\mu K}^2$ only\footnote{The WMAP5 power
spectrum uncertainty at $l=2$ consists of the measurement errors and
the cosmic variance. The later is dominating, but it has nothing to
do with map making or power spectrum estimation uncertainties.
Therefore, we don't have to discuss it here.}. The quadrupole
component ($l=2$) drops for nearly $78\%$ from the WMAP5 value,
whereas the power differences between the WMAP5 and our spectra at
other multiple moments in Fig.~\ref{lowl} are all $<10\%$. It seems
unlikely that a nearly zero power spectrum at $l=2$ can be covered
by the cosmic variance.

\begin{figure*}
\begin{center}
\includegraphics[height=5cm, angle=90]{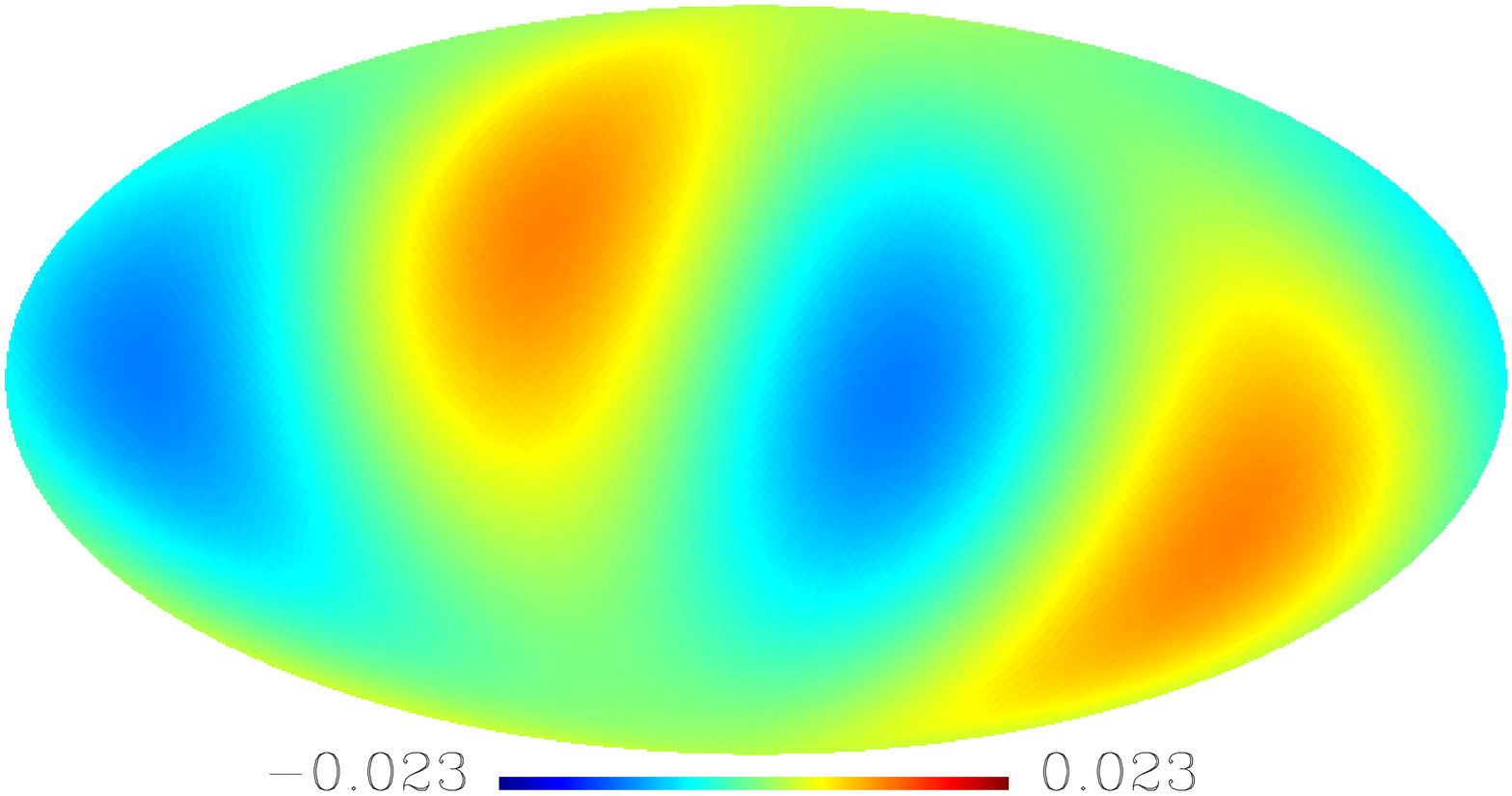}
\hspace{1mm}\includegraphics[height=5cm, angle=90]{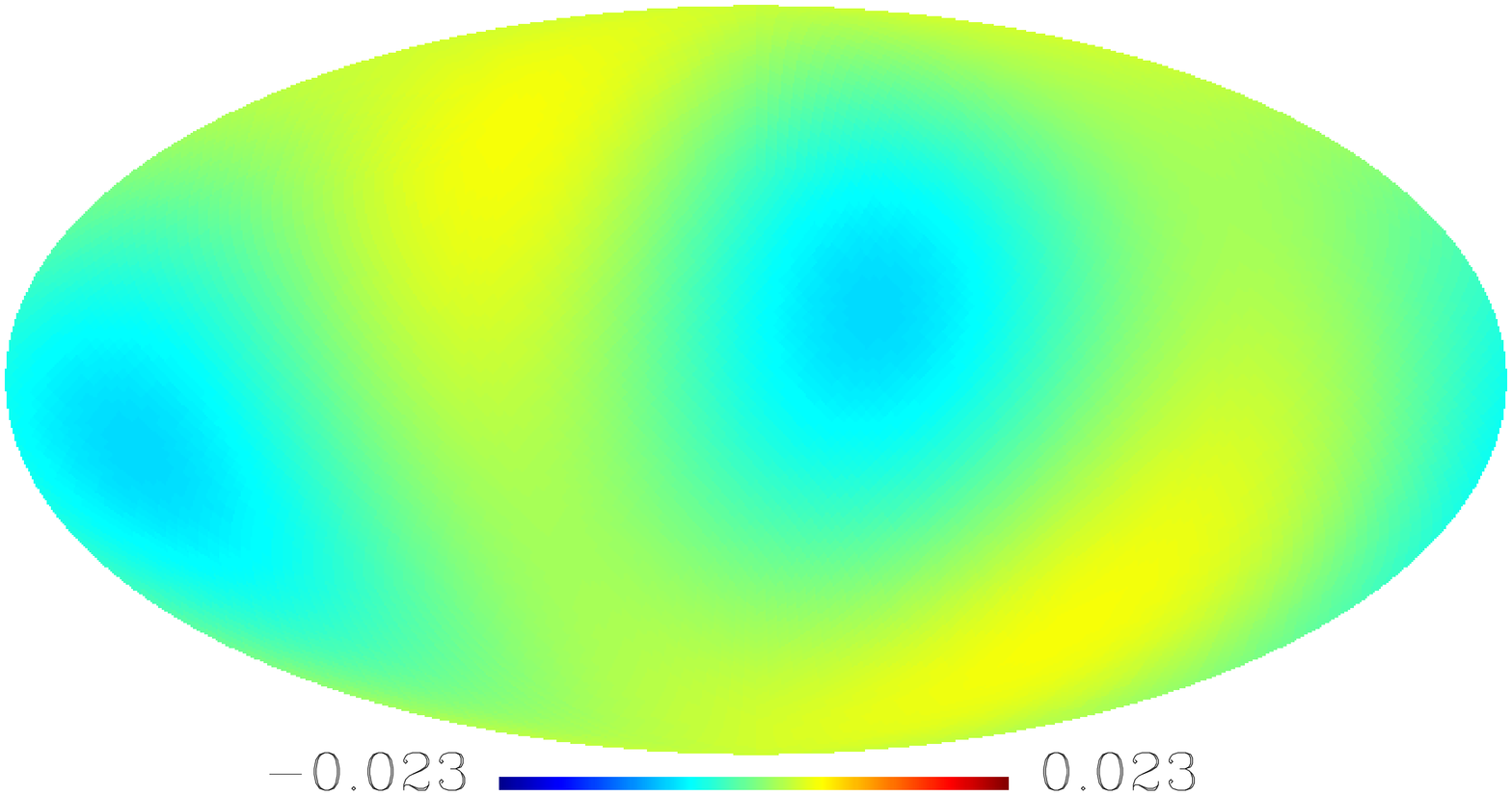}
\hspace{1mm}\includegraphics[height=5cm, angle=90]{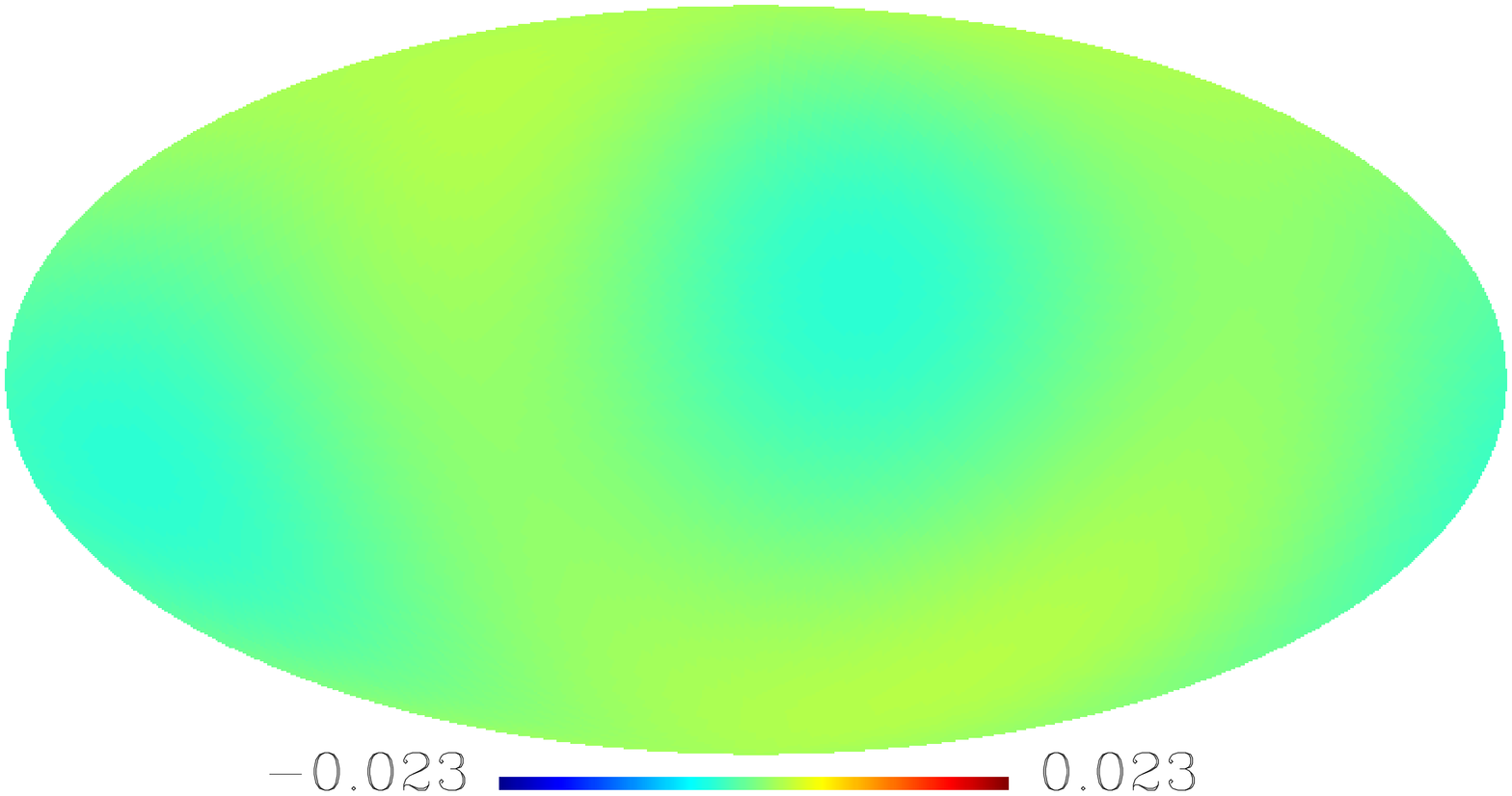}
\includegraphics[height=5cm, angle=90]{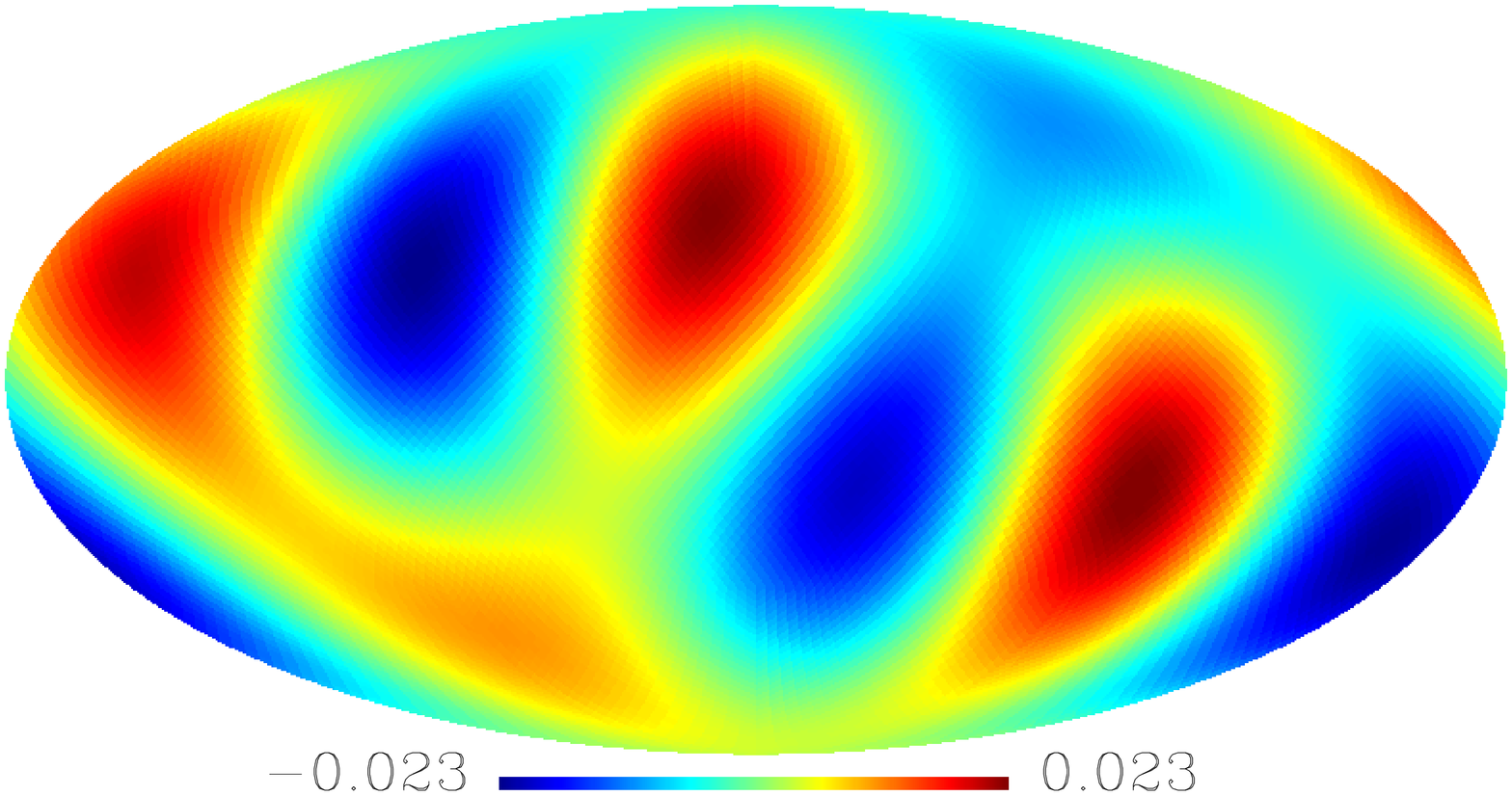}
\hspace{1mm}\includegraphics[height=5cm, angle=90]{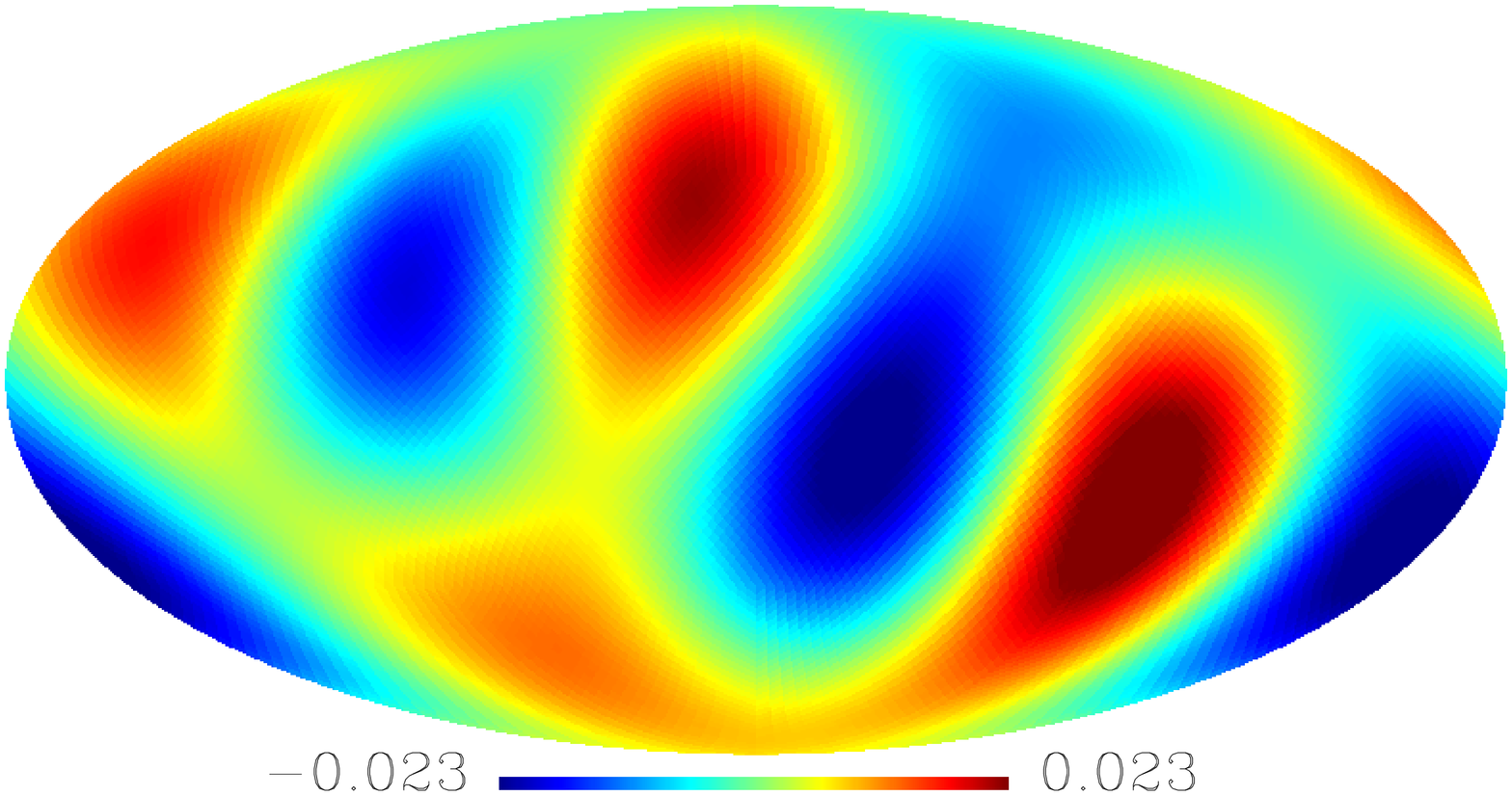}
\hspace{1mm}\includegraphics[height=5cm, angle=90]{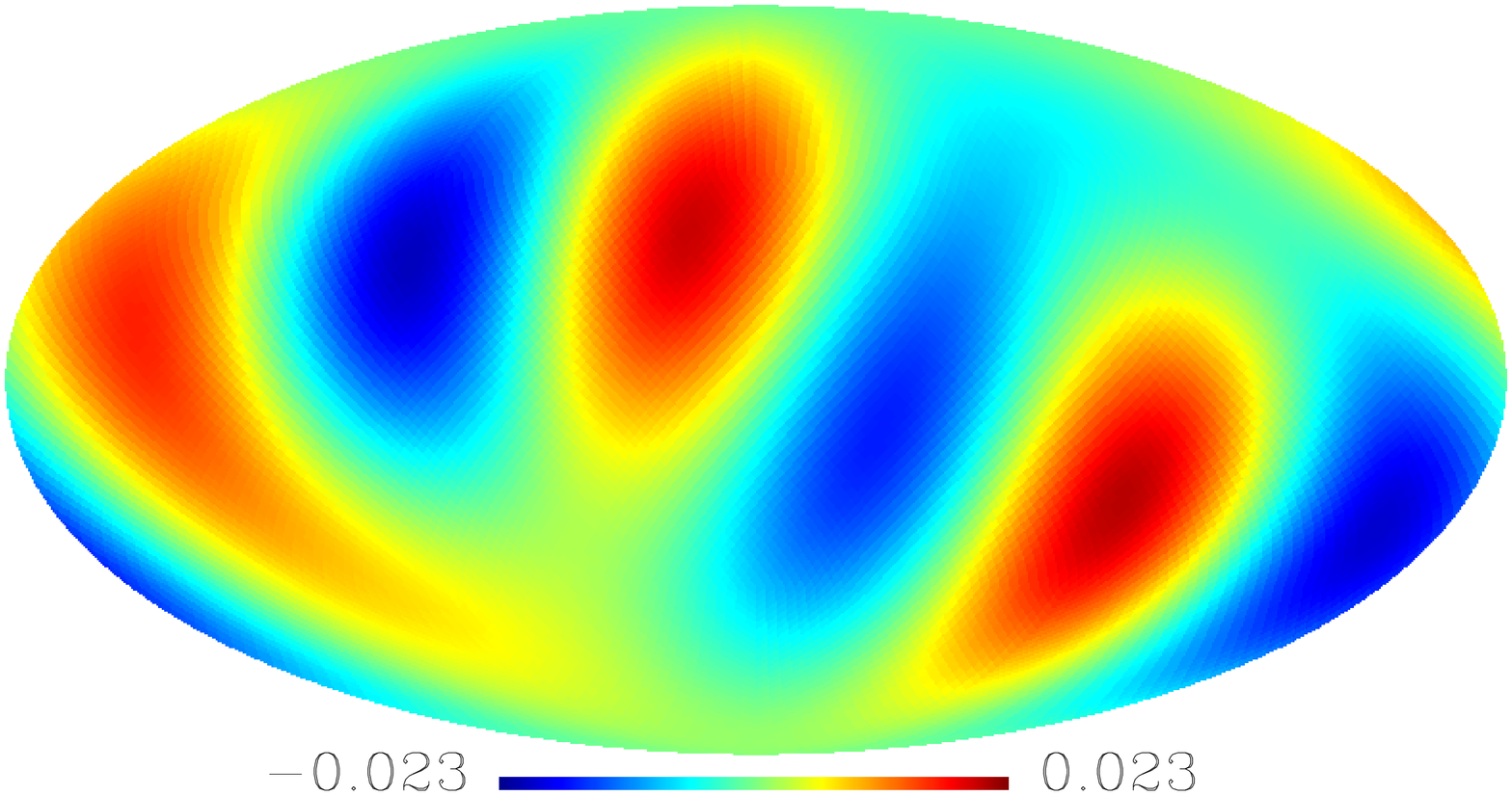}
\end{center}
\vspace{-1mm} \caption{ Quadrupole (upper panel) and octopole (lower
panel) components averaged among 30 V and W bands single-year maps.
From left to right: WMAP5 release, from new maps reconstructed with
 WMAP5 used TOD and
our map-making software, from our safe-mode CMB maps.}
 \label{l2l3}
\end{figure*}

The unexplained orientation of large-scale patterns of CMB maps in
respect to the ecliptic frame has long puzzled cosmologists. By
analyzing CMB maps from the first year WMAP data, Tegmark et al.
(2003) found both the CMB quadrupole ($l=2$) and octopole ($l=3$)
having power along a particular spatial axis, and more works
\cite{cos04,eri04,sch04,jaf05} found that the preferred directions
of these two low-$l$ components are highly aligned and close to the
ecliptic plane.
Recently, significant signal spatially correlated with the ecliptic plane
was detected by Diego et al. (2009) in published WMAP5 map and by Jiang, Lieu
and Zhang (2009) from analyzing spectral variation.
In our maps, the unexplained quadrupole-octopole alignment
disappears: in the WMAP5 release, the angles between the preferred
directions of CMB quadrupole and octopole average out at
$6\degree.5$ for all 30 single year maps; whereas our single year
maps give $26\degree.1$ average angle between the two
components. Fig.~\ref{l2l3} compares the quadrupole and octopole
components averaged over 30 V and W bands our single-year maps
and the corresponding WMAP5 maps results directly. We can see that
both the direction and amplitude of the quadrupole component have
changed significantly from the WMAP5 official maps to our new
maps.

\subsection{\sl High Order Moments}
\begin{table}{Table~1: Average power spectrum ($\rm{\mu K^2}$)}\\[1ex]
\begin{tabular}{l l l l} \hline
 L      & WMAP5 & BOOMRANG\,$^1$& This work\\ \hline
220     & 5716($+2.8\%$)  & 5558(reference) & 5554($-0.1\%$) \\
500-600 & 2454  & 2308 & 2095 \\
800-950 & 2427($+36\%$)  & 1783(reference) & 1766($-1\%$) \\
220-950 & 2584  & 2367 & 2174 \\ \hline 
\end{tabular}
\vspace{6mm}$^1$\,arithmetic average of BOOMRANG98 and 03 
\end{table}

 Fig.~\ref{power spectrum} shows the binned cross power spectra
for high order moments ($l=33$ -- 675), where
the solid line is the released WMAP5 spectrum,
the dashed line the spectrum of new CMB maps from WMAP5 used TOD and
our map-making software, and the dashed-dotted line
the spectrum of safe-mode CMB maps
from WMAP5 safe-mode TOD and our map-making software.
From Fig.~\ref{power spectrum} we see the new
power spectra being systematically lower than the
released WMAP5 power spectrum over multiple moment $l=200$ -- 675.
For example, by averaging all the power spectra in this range, both HEALPix and
PolSpice give our safe-mode power spectrum being averagely $13\%$
lower than WMAP5. Such a power spectrum decrease will cause
important consequences on the best fit cosmological parameters.

It is interesting to see from Table~1 that, in comparison with the
WMAP release, new power spectra in high order moment region are, in
general, better consistent with the BOOMRANG CMB
spectrum~\cite{ruh03,jon06}. Especially for the power spectrum near
the third peak (L=800-950). The relative deviation from the BOOMRANG
spectrum at the first peak ($l=220$) is $\sim 3\%$ for WMAP5 and
only $\sim 0.1\%$ for our safe-mode spectrum respectively; at the
third peak ($l=800-950$), that is $\sim 36\%$ for WMAP5 and only
$\sim 1\%$ for our spectrum, respectively.

\section{COSMOLOGICAL PARAMETERS}
\begin{table*}{Table 2: The best-fit cosmological parameters}\\[1ex]
\begin{tabular}{c c c c c}  \hline
Description & Symbol &  \multicolumn{3}{c}{Value} \\
\cline{3-5}
 & & WMAP5-only\,$^1$ & WMAP5+BAO+SN\,$^1$ & This work \\
\hline
Hubble constant (km/s/Mpc) & $H_0$ & $71.9^{+2.6} _{-2.7}$ & $70.1\pm1.3$ & $71.0\pm 2.7$ \\
Baryon density  & $\Omega_b$  & $0.0441\pm0.0030$ & $0.0462\pm0.0015$ & $0.052 \pm 0.0030$ \\
Dark matter density  &$\Omega_c$&$0.214\pm0.027$& $0.233\pm0.013$& $0.270 \pm 0.027$ \\
Dark energy density &$\Omega_{\Lambda}$ & $0.742\pm0.030$ & $0.721\pm0.015$& $0.678 \pm 0.030$ \\
Fluc. Ampl. at $8h^{-1}$ Mpc & $\sigma_8$ & $0.796\pm0.036$ & $0.817\pm0.026$ & $0.921 \pm 0.036$ \\
Scalar spectral index & $n_s$ & $0.963^{+0.014} _{-0.015}$ & $0.960^{+0.014} _{-0.013}$
& $0.957 \pm 0.015$ \\
Reionization optical depth & $\tau$ & $0.087\pm0.017$ & $0.084\pm
0.016$ & $0.109 \pm 0.017$ \\ \hline \vspace{5mm} $^1$\,{ from
Hinshaw et al. (2008)}
\end{tabular}
\end{table*}

The peak height ratio (first peak$/$second peak, which is closely
related to the baryon density) in the CMB power spectrum is $\sim
2.2$ for WMAP5 release and $\sim2.6$ for our spectrum, indicating
that at least the baryon density should rise. To roughly estimate to
what extent the cosmological parameters can be affected, a brutal
force least-square fit search for the new best-fit parameters is
done in the multiple moment range from 33 to 675
by using the CMBFast package \cite{zal98,zal00,sel96} with the
$\rm{\Lambda CDM}$ model for our new WMAP5 binned power spectrum
and safe-mode spectrum separately. The two sets of results are close to each other.
The results from the safe-mode spectrum are listed in
Table~2. From Fig.~\ref{power spectrum}, we can see the new
power spectrum differs by no more than $20\%$ from the
WMAP5 power spectrum. Therefore, the WMAP5 uncertainties of the
cosmological parameters are roughly suitable for the new
cosmological parameters, as adopted in Table~2.

From Table~2 we can see that, in comparison with the WMAP5 release,
some best-fit cosmological parameters from the improved CMB maps are
significantly different, e.g. the total matter density  $\Omega_m=\Omega_c+\Omega_b$
increases by $\sim 25\%$ (from 0.26 up to 0.32) and the dark energy
density  $\Omega_{\Lambda}$ decreases
by $\sim 10\%$ (from 0.74 down to 0.68).

\section{DISCUSSION}
In this work we observe
remarkably systematic artifacts  in the released WMAP CMB map,
structured signals and noises related with the Solar System, as shown
by Fig.~\ref{fig:residual}, Fig.~\ref{fig:dif-map}
and Fig.~\ref{fig:map}, which turn into the almost all observed
quadrupole component of the released map.
From Fig.~\ref{l2l3} we can see that
the detected ecliptic contamination mostly comes from
the map-making process of the official WMAP temperature map,
the non-zero flagged TOD used
by the WMAP team in their map-making are the secondary source.
It will be very helpful if  the WMAP team can thoroughly recheck
their map-making process to find out where and how the error occurs,
because it is difficult to do so for a non-WMAP team member.

In this work, we built a self-consistent software
package of map-making from TOD and power
spectrum estimation from recovered CMB map. Our software successfully
passes through a variety of tests. With the tested software we reconstruct new CMB maps
from WMAP TOD, which are significantly
different with the official WMAP maps.
Our map-making process prevent the problem
of inconsistency between input TOD and reconstructed map:
the anomaly structure and high fluctuation
observed in the residual TOD of official WMAP maps disappear in the residual TOD
 of our new maps.
Therefore, the new CMB maps are certainly improved by our map-making
and more reliable than the WMAP release.

The difference between the old and new
temperature maps is almost the same with
the quadrupole component of the released WMAP5 CMB map (Fig.~\ref{fig:map}).
Being consistent with this, we
also see a structure  similar to the released WMAP5 quadrupole component
after smoothing the map of the
WMAP residual TOD (Fig.~\ref{fig:residual}) to $N_{side}=8$.
The correlation map of the residuals between input TOD and predicted by
the released WMAP temperature map is also very similar to the quadrupole component
observed in the released WMAP CMB map.
Such phenomenon strongly suggests that, in the WMAP5 CMB temperature map,
a very large portion of the presently known quadrupole component
is actually artificial.
As a natural consequence, in our new CMB map the $l=2$ fluctuation
falling near the ecliptic plane no longer exists,
and then the quadrupole component is nearly zero.

Despite cosmic variance, the anisotropy at the lowest harmonics as inferred
from our reprocessed sky maps are likely to be too low to accommodate the standard
 cosmological parameters.
A nearly zero quadrupole component of CMB map places a tight constraint on the nature
of the early universe.
As well known, the subtended angle of the cosmological horizon at last scattering
is only about 1 - 2 degree.
Measured CMB power at the large angular scale of
lowest moments should reflect the circumstance of the inflation period,
e.g. the state of matter and the production of density fluctuations
during inflation. Our result indicates that we may catch sight of
the very early rigid universe, the primordial density perturbations
should be generated in a later stage of inflation.
It is hopeful to study the generation and revolution of density fluctuation
through more reliable large scale CMB data and with suitable mathematical
tool for data analysis in the future.

Besides the ecliptic foreground contamination revealed in this work,
we have found other two kinds of systematic distortions in official WMAP maps.
One is the Galactic foreground contamination.
Although the WMAP team has paid huge attention to separate the temperature
anisotropy of CMB from Galactic foreground emission,
in released WMAP CMB maps we still find notable distortion from
hot Galactic sources~\cite{liu09}: pixels $141\degree$ away from the hottest pixels
in the Galactic plane are on average 12 -- 14 $\rm{\mu K}$ cooler than average pixels.
Another one is the no-negligible effect of imbalance WMAP observations~\cite{li09}:
significant correlation between the map temperature released by the WMAP team
and the difference in the observation numbers that a pixel is pointed by the satellite's
two antennae,
the temperature distortion caused by this observational effect is as high as up to
$\sim 20\,\rm{\mu K}$.
These two kinds of systematics are also detected in our new maps
and should be further inspected and corrected. Therefore
we believe that there still exist space to improve
CMB maps and accuracies of cosmological parameters again by further
improving map-making from WMAP TOD data and removing residual systematic
errors in recovered temperature maps.

 Except the quadrupole component, the power differences between the
WMAP5 and our  low-$l$ spectra at other multiple moments
(Fig.~\ref{lowl}) are not too large. The anomalously planar cosmic octopole
still remains in our new WMAP5 CMB map, and also in the safe-mode map
just with a little lower amplitude (see Fig.~\ref{lowl} and Fig.~\ref{l2l3}).
However, from our recent analysis, we find that the effect of imbalance WMAP observations
may contribute to some extent to the observed octopole component in WMAP CMB maps,
and the real spectrum at $l=3$ might be lower than observed.
It is expected that more reliable low-$l$ spectrum can be derived
after better understanding and properly correcting systematics in
observation and problems in map-making. A reliable large-scale spectrum
will certainly help to understand the early universe.
For example, for the primordial density perturbations, inflation theory predicts
 the index $n$ in the scalar spectrum  $P(k)\sim k^n$ being nearly unit,
$n\approx 1$,
over a very wide range of $k$ by virtue of scale invariance: most perturbations
exited the horizon with the same amplitude in the brief interval
of rapid exponential expansion.
There is now a possibility that a larger $n$ at low $l$
might be more consistent with the data.
This in turn would mean that quantum fluctuations generated during inflation
may no longer be a viable source of super-horizon perturbations without
significant modification.

The best fit cosmological parameters of this work shown in Table~2
are just preliminary results. It is interesting that, for nearly all the basic cosmological
parameters ($H_0$, $\Omega_b$, $\Omega_c$, $\Omega_{\Lambda}$,
$\sigma_8$, $n_s$), both our values and the WMAP + BAO + SN
cosmological parameter values ~\cite{hin08} are simultaneously
higher or lower than the WMAP5 alone ~\cite{hin08}. This shows that
the new cosmological parameters given by us (although still rough)
match with improved accuracy those of  other two
different kinds of measurements: the distance measurements from the
Type Ia supernovae (SN) and the Baryon Acoustic Oscillations (BAO)
in the distribution of galaxies.

\acknowledgments Profs. Richard Lieu, Tan Lu, Miao Li,
Xinmin Zhang, Charling Tao and the anonymous referee
are thanked for helpful comments and suggestions.
This work is supported by the National Natural
Science Foundation of China (Grant No. 10533020) and
the National Basic Research Program of China (Grant No. 2009CB-824800).
 The data analysis made use of the WMAP data archive and the
HEALPix, PolSpice, CMBFast software packages.

\label{lastpage}
\end{document}